\documentclass[10pt,twocolumn,letterpaper]{article}
\pdfoutput=1

\usepackage{iccv}
\usepackage{times}
\usepackage{epsfig}
\usepackage{graphicx}
\usepackage{amsmath}
\usepackage{amssymb}
\usepackage{booktabs}
\usepackage{makecell}
\usepackage{enumitem}
\usepackage{titling}
\date{}

\usepackage[pagebackref=true,breaklinks=true,letterpaper=true,colorlinks,bookmarks=false]{hyperref}

\iccvfinalcopy 

\def\iccvPaperID{10452} 

\ificcvfinal\fi

\begin{document}

\title{Time-Multiplexed Coded Aperture Imaging: Learned Coded Aperture and Pixel Exposures for Compressive Imaging Systems}

\author{Edwin Vargas$^{1,*}$, Julien N.P. Martel$^{2,*}$, Gordon Wetzstein$^2$, Henry Arguello$^1$\\
$^1$Universidad Industrial de Santander, Colombia $^2$Stanford University, USA\\
    {\tt\small edwin.vargas4@correo.uis.edu.co, jnmartel@stanford.edu} \\
    {\tt\small gordon.wetzstein@stanford.edu, henarfu@uis.edu.co}
}

\maketitle
\ificcvfinal\thispagestyle{empty}\fi

\begin{abstract}
Compressive imaging using coded apertures (CA) is a powerful technique that can be used to recover depth, light fields, hyperspectral images and other quantities from a single snapshot. 
The performance of compressive imaging systems based on CAs mostly depends on two factors: the properties of the mask's attenuation pattern, that we refer to as ``codification" and the computational techniques used to recover the quantity of interest from the coded snapshot.
In this work, we introduce the idea of using time-varying CAs synchronized with spatially varying pixel shutters. We divide the exposure of a sensor into sub-exposures at the beginning of which the CA mask changes and at which the sensor's pixels are simultaneously and individually switched ``on" or ``off". 
This is a practically appealing codification as it does not introduce additional optical components other than the already present CA but uses a change in the pixel shutter that can be easily realized electronically. 
We show that our proposed time multiplexed coded aperture (TMCA) can be optimized end-to-end and induces better coded snapshots enabling superior reconstructions in two different applications: compressive light field imaging and hyperspectral imaging. 
We demonstrate both in simulation and on real captures (taken with prototypes we built) that this codification outperforms the state-of-the-art compressive imaging systems by more than $4$dB in those applications.
\end{abstract}
\vspace{-1em}
\section{Introduction}
\label{sec:intro}
\begin{figure}[t!]
	\centering
	\includegraphics[width=0.99\linewidth]{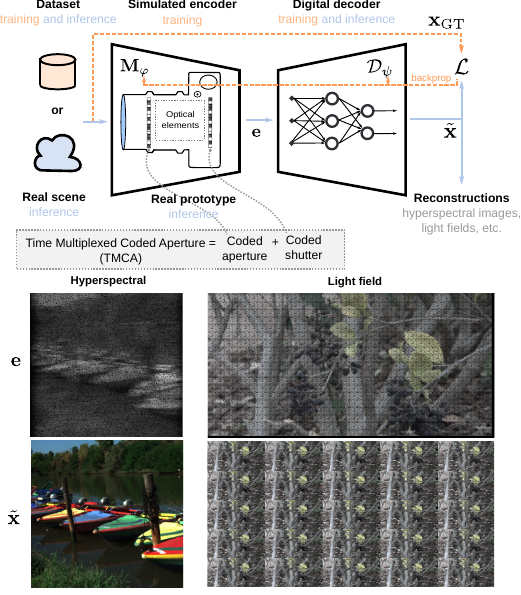}
	\caption{\small An illustration of the proposed Time Multiplexed Coded Aperture (TMCA) codification in the proposed end-to-end differentiable pipeline. We also show coded snapshot and reconstruction examples for our compressive light field and hyperspectral imaging applications.}
	\vspace{-2em}
	\label{fig:teaser}
\end{figure}
\let\thefootnote\relax\footnotetext{$^*$ denotes equal contributions.}
Computational imaging techniques combining the co-design of hardware and algorithms have successfully enabled the development of novel cameras for several applications such as spectral imaging \cite{arce2013compressive}, depth imaging \cite{chang2019deep}, light field imaging \cite{marwah2013compressive}, or computed tomography \cite{kalender2006x}. Among those, in particular compressive imaging \cite{donoho2006compressed} approaches that aim at multiplexing visual information in a single snapshot have been popularized to reconstruct high dynamic range images, videos, spectral or depth information \cite{metzler2019deep,yuan2021snapshot,SCCSI_Correa}. They use optical and electronic coding strategies to encode light and use physics-based forward models along with optimization techniques to reconstruct the aforementioned visual quantities. Besides their ability to recover visual information not easily accessible with traditional cameras, another advantage of compressive imaging systems is they can realize sparse measurements, for instance capturing a single coded snapshot to recover a whole light-field. This yields typically ill-posed reconstruction problems that can nevertheless be solved by leveraging knowledge about the signal's sparsity and using other priors about the visual quantity to reconstruct. For optimal reconstruction, compressive sensing theory usually assumes dense uncorrelated measurement matrices. However, this is rarely the case, as those measurement matrices are induced by design constraints and the physical realization of the optical and electronic codifications. As a result, those matrices are almost always sparse and highly structured, thus impeding the performance of compressive imaging systems. Hence, the main lever to improve those systems is to improve their codification.

In particular, many compressive imaging systems employ coded aperture (CA) masks as codification elements \cite{raskar2006coded, levin2007image,  wagadarikar2008, sankaranarayanan2010compressive, marwah2013compressive,arguello2014colored, asif2016flatcam}. CAs are physical components inserted in the optical system to spatially modulate light intensity. As an example, the coded aperture snapshot spectral imaging system (CASSI) presented in \cite{wagadarikar2008single,arce2013compressive} employs a dispersive element behind a CA used to select the spatial locations that get spectrally dispersed. Those CAs can be fabricated using inexpensive technologies (such as chrome-on-quartz) for simple binary codes. However creating multi-valued color codes, yielding much better reconstructions, requires expensive microlithography and coating technologies \cite{arguello2014colored}. Yet another example illustrating the use of CA masks in compressive imaging is in the reconstruction of light fields, for instance by using a CA inserted between the sensor and the objective lens \cite{marwah2013compressive}. In this system, correlations in the measurement matrix are created by the similarity between the angular views and typically limit the quality of the light-field reconstructions. These two examples illustrate the need for CA systems that yield better codifications and that are still easy and inexpensive to implement.

This work addresses this challenge by proposing a new codification we call Time-Multiplexed Coded Aperture (TMCA). It induces a better conditioning of those measurement matrices and can be realized using existing components that are widely used in compressive imaging systems. The proposed TMCA consists of using a conventional amplitude CA along with a coded exposure. Coded exposures are temporal modulations of the pixel of a sensor. In a coded exposure, a shutter function turns individual pixels ``on" or ``off" during the exposure, thus modulating light integration. Pixel-wise shutters can be realized electronically modifying the pixel architecture \cite{martel2020neural,wei2018coded,luo2019cmos,sonoda2016high} or using spatial light modulators (SLMs) \cite{hitomi2011video}. More specifically, we generate a TMCA by synchronizing a CA that changes its pattern in time --dubbed as a time-varying CA-- and a spatially varying shutter-function realizing a coded exposure. We show that this combination produces a new family of CAs with better codification capabilities that we demonstrate in two specific compressive imaging applications: hyperspectral imaging and light field imaging. 

Furthermore, we propose to learn the TMCA codification, inspired by works in deep optics  \cite{martel2020neural,chang2019deep,metzler2019deep,wu2019phasecam3d,nehme2020deepstorm3d}. We perform the joint optimization of the TMCA codes along with a neural network (NN) used to recover the light fields or hyperspectral images. Our end-to-end differentiable approach can be interpreted as an encoder-decoder framework in which the encoder performs the optical-electronic codification and a digital decoder (the NN) is used for reconstruction. The CA and shutter functions are optimized for each application in simulation, considering the specific constraints of the spatial light modulators realizing the CA and of the sensor realizing the coded-exposures. After training, the optimized TMCA codes can be deployed to physical devices that can be used to capture real-world scenes. 

We summarize the contributions of our work as follows:
\begin{itemize}[leftmargin=*] \itemsep0em 
\item We introduce a new codification for compressive imaging systems called time-multiplexed coded aperture (TMCA). 
\item We develop new forward models based on the use of TMCAs for two applications: hyperspectral imaging and compressive light field imaging. In the first case, we show TMCA emulates an expensive color filter array, while in the second, the TMCA is an angular sensitive CA resulting in less correlations in the coding of angular views.
\item We learn, in simulation, differentiable optical electronic encoders realizing the TMCAs as well as a NN decoders solving the reconstruction problems. We demonstrate that the learned codes compare favorably against baselines that use traditional CA as well as against our own TMCA baselines using non-optimized codes.
\item We build two prototypes: a compressive light field imaging system and a hyperspectral imager. We translate the learned TMCA codes to hardware and conduct real experiments showing the better results of TMCA in simulation transposes to real-world systems.
\end{itemize}

\section{Related work}
\label{sec:related}
\paragraph*{Coded aperture systems} employ carefully designed mask patterns to encode incident light. They can be thought of as arrays of pinholes that were developed to improve upon the light efficiency of single pinhole cameras. CA-based systems are widely employed in astronomy or biomedical applications in which lenses cannot easily be fabricated because of the wavelengths at play \cite{fenimore1978coded,gottesman1989new}. Coded snapshots captured through CAs can be decoded using computational techniques to provide sharp, clean images. Recent works have considered CA methods for developing novel image acquisition techniques for depth imaging \cite{levin2007image}, motion deblurring \cite{raskar2006coded}, lensless imaging \cite{zomet2006lensless,asif2016flatcam}, high-dynamic range \cite{nayar2003adaptive}, video imaging \cite{marcia2009compressive}. Furthermore, compressive sensing  methods  have  also  been used  jointly  with  CAs: examples include spectral  imaging \cite{wagadarikar2008single, arguello2014colored},  dual-photography \cite{sen2009compressive}, light field imaging \cite{marwah2013compressive}, or image super-resolution \cite{mohan2008sensing}. The quality of the reconstruction in all these applications mainly depends on the codification created by the CA. Hence, our work could readily be adapted for any of those applications and we believe, would directly improve them. Here we choose to focus on hyperspectral imaging and light field imaging.
\vspace{-1em}
\paragraph*{Hyperspectral imaging} (HSI) aims at capturing images with a large number of spectral channels (more than the three typical red, green, blue bands) \cite{kim20133d}. There are three main approaches for HSI: computed tomography imaging, spectral scanning, and snapshot compressive imaging. Based on a dispersive optical element, such as a prism or a diffraction grating, scanning-based approaches can capture each wavelength of light in isolation through a slit: so-called whiskbroom or pushbroom scanners \cite{brusco2006system,porter1987system}. While scanning methods yield high spatial and spectral resolution, they are typically slower than other methods. Computed tomography imaging spectrometry \cite{habel2012practical,johnson2007snapshot,okamoto1993simultaneous} was introduced to mitigate this limitation. It employs a diffraction grating that difracts incident collimated light into patterns in different directions, even though such systems can be real-time, this is at the expense of spatial resolution. Finally coded aperture snapshot spectral imaging (CASSI) methods \cite{wagadarikar2008,SCCSI_Correa,baek2017compact} were also introduced for faster captures. Similar to other compressive imaging techniques, those are limited by the codification properties as well the reconstruction algorithms they use. Our work addresses those limitations introducing a new coding strategy and learning the codes in an end-to-end fashion. 
\vspace{-1em}
\paragraph*{Light field imaging} aims at capturing the amount of light passing through every direction in any point in space, practically representing a scene through different ``angular perspectives". Early light-field (LF) camera prototypes either used a film sensor using pinholes or microlens arrays \cite{ives1903parallax}. More recently, camera arrays \cite{wetzstein2013plenoptic,wilburn2005high,levoy1996light,liang2008programmable} improved the quality of LF captures. However, these approaches may require multiple cameras and snapshots and are often impractical or expensive to build. Using compressive sensing, LF architectures have been proposed with the goal of taking fewer snapshots \cite{ashok2010compressive,babacan2012compressive,marwah2013compressive}. Marwah et al. \cite{marwah2013compressive} first introduced coded light field photography (CLFP), in which a light field can be recovered from a single coded measurement obtained with a CA inserted between the sensor and the objective lens. Hirsh et al. \cite{hirsch2014switchable} proposes another coding solution using angular sensitive pixels (ASP) that has been shown to improve the conditioning of the measurement matrix over CA approaches such as \cite{marwah2013compressive}. A limitation of this approach is that it uses specific sensors with ASP of given angular and frequency responses. Our work allows the generation of analogous codifications using a CA, improving the reconstruction but using more flexible codifications (they can be changed) and hardware.
\vspace{-1em}
\paragraph*{End-to-end optimization} and the co-design of optics and algorithms is at the core of computational photography. Automatic differentiation programming tools \cite{paszke2017automatic,abadi2016tensorflow}, enable the implementation of fully differentiable pipelines and have fueled this concept with a number of applications such as color imaging and demosaicing \cite{chakrabarti2016learning}, extended depth of field imaging \cite{sitzmann2018end}, depth imaging \cite{chang2019deep,haim2018depth,wu2019phasecam3d}, image classification \cite{chang2018hybrid, muthumbi2019learned}, HDR imaging \cite{metzler2019deep}, microscopy \cite{nehme2020deepstorm3d, kellman2019data}. Our work builds on end-to-end optimization techniques, in the line of \cite{martel2020neural}, it does not only optimize the optical encoding but a part of the sensor itself: the coded exposures, together with the NN used for reconstruction.

\section{Time-multiplexed coded apertures}
\label{sec:methods}
\begin{figure}[t!]
	\centering
	\includegraphics[width=\linewidth]{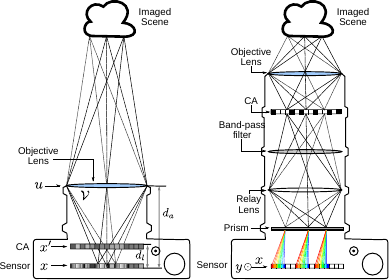}
	\caption{\small A diagram of the ray optics for our light field imaging (left) and hyperspectral imaging (right) systems using TMCAs.}
	\label{fig:cs_systems}
\end{figure}
The principle of the proposed TMCA consists of synchronizing a time-varying CA with a shutter function realizing a coded exposure. This idea can be used to realize different systems depending on the specificities of the optical setup it is coupled with, i.e the optical elements in between the CA and the sensor as shown in Figure~\ref{fig:teaser}.
First, we present the general forward model and codification induced by a TMCA. We then describe how this model can be used for two specific applications using different optical setups: in light field and hyperspectral imaging, derive the codification for those two systems and demonstrate how the proposed TMCA improves the traditional CAs they use.
\newcommand{\dd}{\,\mathrm{d}}
\subsection{Generalities}
\label{subsec:tmca_generalities}
We consider the irradiance $I$, invariant in time, the quantity we reconstruct with our compressive imaging systems. Our proposed TMCA consists of two coding stages.

The first stage optically encodes $I$ into a field $g$ incident to the sensor, using a time-varying CA $T(t)$. We model $g(t)$ as the response of a linear optical system $\mathcal{O}$ given input $I$:
\begin{equation}
    g(t) = \mathcal{O}(T(t),I).
\end{equation}
As we shall see in section~\ref{subsec:tmca_lf}~and~\ref{subsec:tmca_hs}, the exact form of $\mathcal{O}$ is application dependent but is always a function of the time-varying coded exposure $T(t)$. 

The second stage consists of a coded exposure. The irradiance $g(t)$ is captured by the sensor in a single snapshot of exposure time $\Delta{t}$. During this exposure, a spatially-varying shutter function $S_{i,j}(t)$ turns the pixel $(i,j)$ ``on" and ``off" multiple times, resulting in the coded exposure $e_{i,j}(t)$ in which the integration of $g(t)$ has been modulated:
\begin{equation}
e_{i,j}(t) = \int_{{t}}^{t+\Delta t} S_{i,j}(t')g_{i,j}(t')\dd t'.
\label{eq:exposure_coded}
\end{equation}

In particular, we consider binary shutter functions $S_{i,j}$ defined on $K$ discrete time slots (sub-exposures of time $\delta{t}$): we have $\Delta{t}=K\cdot\delta{t}$. The coded exposure can then be rewritten as
\begin{equation}
\label{eq:exposure_coded_discrete}
e_{i,j} = \sum_{k=0}^{K-1} S_{i,j}^k g_{i,j}^k, \text{with } S_{i,j}^k \in \{0,1\}^k,
\end{equation}
where $g_{i,j}^k$ and $S_{i,j}^k$ denote the irradiance incident on the sensor and shutter function at pixel $(i,j)$ in the $k-$th time slot. 
Those slots allow us to synchronize the coded apertures with the coded exposures (both are indexed by $k$).

Another way to write this discrete model is using the matrix-vector notation $\mathbf{e} = \sum_k \mathbf{\boldsymbol{S}}^k\mathbf{g}^k$, where $\mathbf{e}$ and $\mathbf{g}$ are the ``vectorized" form of the exposure and coded irradiance and $\boldsymbol{S}$ is the measurement matrix representing the shutter function. The irradiance incident to the sensor is $\mathbf{g}^k = \boldsymbol{O}^k \mathbf{x}$, where the matrix $\boldsymbol{O}^k$ is the point spread function of the application-dependent optical system, also including the coded aperture, and $\mathbf{x}$ represents the irradiance $I$ in its vector form. Using this notation, the forward model of the proposed TMCA can be simply written as
\begin{equation}
\label{eq:exposure_coded_matrix}
\mathbf{e} = \sum_{k=0}^{K-1} \boldsymbol{S}^k\boldsymbol{O}^k \mathbf{x} = \bold{M}\bold{x},
\end{equation}
defining the overall measurement matrix of our compressive imaging system $\mathbf{M}= \sum_{k=0}^{K-1} \boldsymbol{S}^k \boldsymbol{O}^k$. Recovering $\bold{x}$ given the coded snapshot $\bold{e}$ amounts to solving an inverse problem.

In the following sections, we use this same general model for two different applications. In our first application of compressive light field imaging, we aim at recovering light fields and the irradiance $I$ (and thus $\bold{x}$) considers multiple view angles. Our second application targets hyperspectral imaging. In that case, $I$ (and $\bold{x}$) considers multiple frequency bands.

\subsection{TMCA for compressive light field imaging}
\label{subsec:tmca_lf}
We consider the same optical setup proposed for compressive light field imaging by Marwah et al. \cite{marwah2013compressive}. A coded aperture mask $T$ is placed between the objective lens and the sensor, at a distance $d_l$ from the latter. The incident field $g(x,t)$ at the sensor is the spatially modulated light field projected along its angular dimension taken over the aperture area $\mathcal{V}$:
\begin{equation}
\label{eq:LFCS}
g(x) = \int_{\mathcal{V}} l(x,u)\,T\big(x + s(u-x)\big) \dd{u},
\end{equation}
where $s=d_l/d_a$, with $d_a$ the distance from the sensor to the aperture plane, is the shear of the mask pattern with respect to the incident light field $l(x,u)$. Similar to \cite{levoy1996light}, we adopt a two-plane parameterization for the light field where $x$ is the 2D spatial dimension on the sensor plane and $u$ denotes the 2D position on the aperture plane (see Fig. \ref{fig:cs_systems}).
\begin{figure}[t!]
	\centering
	\includegraphics[width=\linewidth]{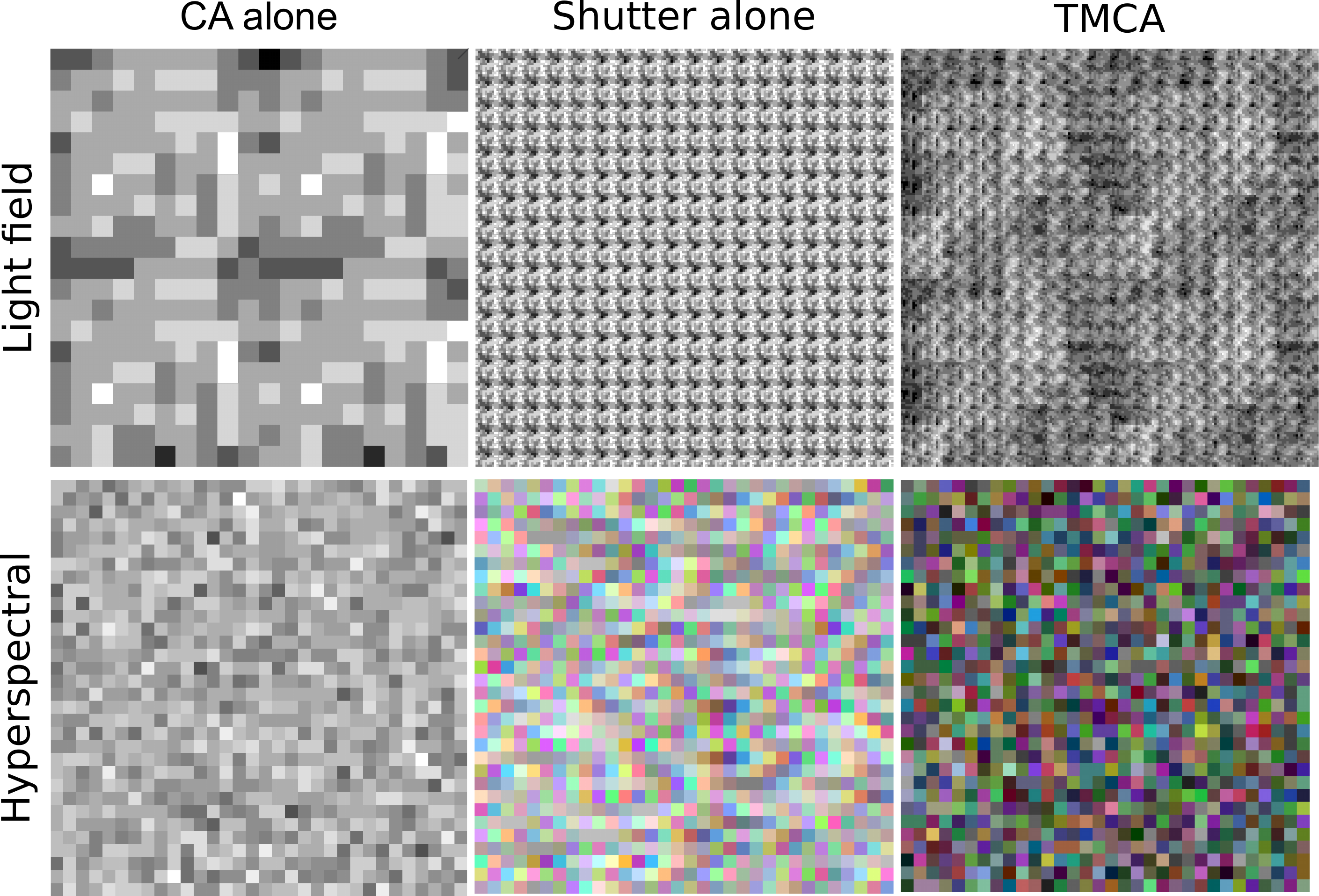}
	\caption{\small A plot illustrating the codes for our two applications for three different cases of codification: ``CA only", ``Shutter only" and the synchronization of both yielding the proposed TMCA.\vspace{-1.5em}}
	\label{fig:tmca_codes}
\end{figure}
Now considering the time-varying coded aperture $T$ and the shutter function $S$, introduced in Section~\ref{subsec:tmca_generalities}, the measurement model yielding the exposure $e$ is
\begin{equation}
\label{eq:LF_prop_cont}
e(x) = \int_{\mathcal{V}} \int_{\Delta{t}} S(x,t')\,l(x,u)\,T(x + s(u-x),t') \dd{u} \dd{t'}.
\end{equation}
By defining the TMCA $\hat{T}$ as 
\begin{equation}
\label{eq:TMCA_LF_sensor}
\hat{T}{}(x,u) = \int_{\Delta{t}} S(x,t')\,T(x + s(u-x),t') \dd{t'},
\end{equation}
the model in Equation~\eqref{eq:LF_prop_cont} can be simply rewritten as
\begin{equation}
\label{eq:LF_prop_cont_2}
e(x) = \int_{\mathcal{V}} l(x,u)\,\hat{T}(x,u) \dd{u}.
\end{equation}

The model from \cite{marwah2013compressive} described in Equation~\eqref{eq:LFCS} does not include the coded-exposure. Comparing the latter with our TMCA model in Equation~\eqref{eq:TMCA_LF_sensor}, we note that our model can be seen as an equivalent coded aperture we denoted $\hat{T}$.

Furthermore, using a change of variable, we can show that the proposed TMCA for compressive light field imaging can be expressed in the coded aperture plane as
\begin{equation}
\label{eq:TMCA_LF}
\hat{T}(x',u) = \int_{\Delta t} S(x' + \hat{s}(u-x'),t')T(x',t') \dd{t'},
\end{equation}
using the spatial coordinates $x'$ in the coded aperture plane (see Fig.\ref{fig:cs_systems}) and with $\hat{s}=d_\ell/(d_a-d_\ell)$. 

Interestingly, this shows that using the proposed TMCA, the equivalent CA's pixels can be seen as responding differently to rays coming from different angles. If the shutter function is removed and the CA remains constant in time, then the TMCA reduces to \cite{marwah2013compressive}, where all the pixels respond equally for all angles. If we were now to consider the shutter function alone, without the CA, the time-modulation in the sensor plane averages all the views, which brings no clear coding advantage if the scene is static in time (since all the views would be the same). In the top row of Figure~\ref{fig:tmca_codes} we plot the codification in the sensor plane.
\subsection{TMCA for compressive hyperspectral imaging}
\label{subsec:tmca_hs}
%
%
We use an optical design similar to the coded aperture spectral snapshot imager (CASSI) proposed in~\cite{wagadarikar2008single} for hyperspectral compressive imaging. In this architecture, spectral dispersion is achieved using a prism between the lens and the sensor. The quantity we aim at recovering is the irradiance $I(x,y,\lambda)$. Note that we are now explicitly considering a second spatial dimension $y$ (and will assume the prism disperses in the $x$ dimension) and the spectral dimension $\lambda$. Similarly, we denote the spatial dependency of the coded aperture $T(x,y)$. The field impinging the sensor is now also dependent on the optical response of the prism $h$ as well the spectral response of the sensor $\kappa$:
\begin{multline}
\label{eq:CSI_cont}
g(x,y) = \iiint T(x',y')\,I(x',y',\lambda)\\ 
h(x-\mathfrak{s}(\lambda) - x',y-y')\,\kappa(\lambda)\dd{x'}\dd{y'}\dd{\lambda},
\end{multline}
where $\mathfrak{s}(\lambda)$ is the wavelength dependent spatial shift induced by the prism.

Using the shutter function $S$ to create the TMCA, the optically encoded field $g$ yields the coded-exposure
\begin{multline}
\label{eq:CSI_prop_cont}
e(x,y) = \int_{\Delta t} S(x,y,t')\iiint T(x',y',t')\,I(x',y',\lambda)\\ 
h(x-\mathfrak{s}(\lambda) - x',y-y')\,\kappa(\lambda) \dd{x'}\dd{y'}\dd{\lambda}\dd{t'}.
\end{multline}
Since $h$ is the propagation through unit magnification imaging optics and a dispersive element with linear dispersion, the impulse response can be expressed as $h(x-\mathfrak{s}(\lambda) - x',y-y')=\delta(x-\lambda - x',y-y')$. After substituting this expression in \eqref{eq:CSI_prop_cont} simplifying and rearranging (see supplemental), we can express the coded measurements as 
\begin{multline}
\label{eq:CSI_prop_cont_4}
e(x,y) =  \iiint \hat{T}(x',y',\lambda)\,I(x',y',\lambda)\\ 
\delta(x-\lambda - x',y-y')\,\kappa(\lambda) \dd{x'}\dd{y'}\dd{\lambda},
\end{multline}
that use the TMCA $\hat{T}$ defined as
\begin{align}
\hat{T}(x',y',\lambda)=\int_{\Delta t} S(x'+\lambda,y',t')T(x',y',t') \dd{t'}.
\label{eq:TMCA_S}
\end{align}
\begin{figure}[t!]
	\centering
	\includegraphics[width=0.94\linewidth]{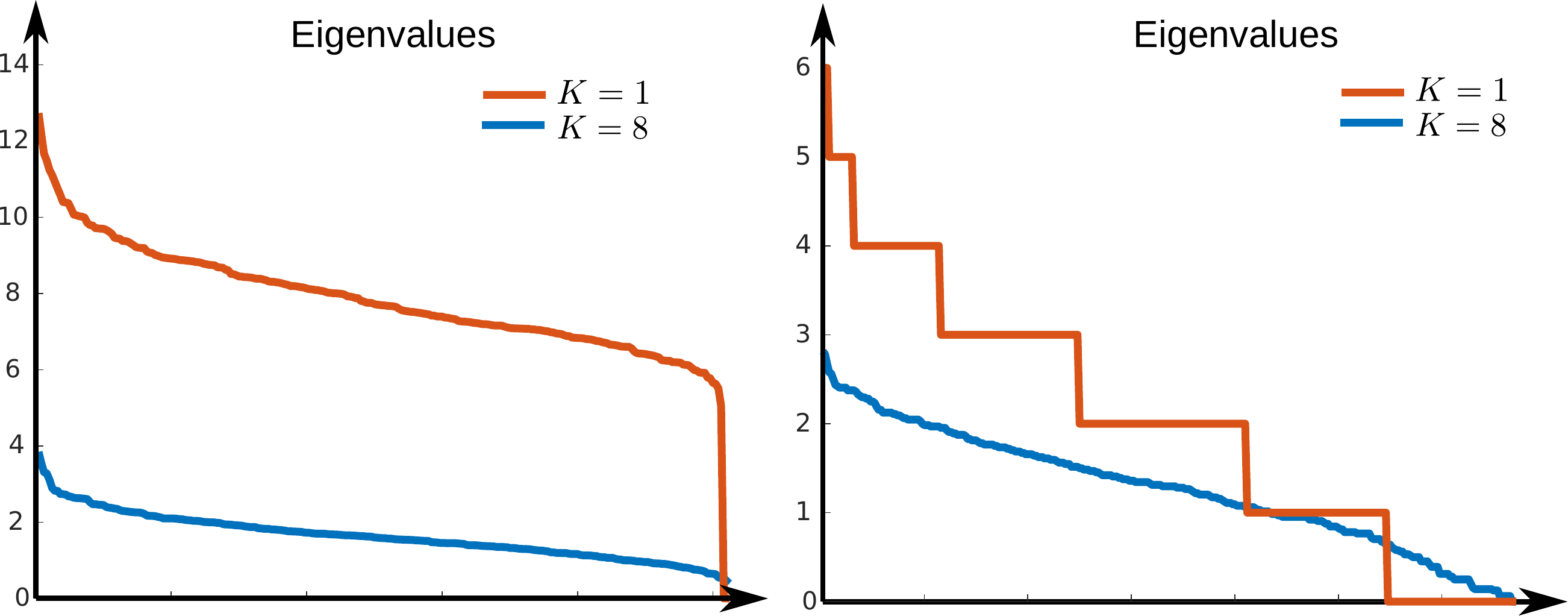}
	\caption{\small Eigenvalues distribution sorted in descending order for the compressive light field (left) and compressive spectral imaging system (right) for the case of $K=1$ (traditional, no coded exposure) compared with the case of $K=8$ (more integration slots in the TMCA).\vspace{-1.5em}}
	\label{fig:eigenvalues}
\end{figure}
In summary, adding the coded exposure to the model of \cite{wagadarikar2008single} generates a TMCA with a new coded aperture $\hat{T}$. The dependency of Equation~\eqref{eq:TMCA_S} in $\lambda$ shows the proposed TMCA emulates a color coded aperture which would otherwise be expensive to create.
Importantly, note this is only true when both the shutter function and CA are jointly employed. If the shutter function is a constant, this is exactly the model in \cite{wagadarikar2008single}: there is no spectral response of the CA. On the other hand, if the CA is removed, $\hat{T}$ exhibits a spectral response that shares the same code for every wavelength that is simply shifted in the $x$ direction depending on the wavelength.

A final advantage of the TMCA codification is that even in the case we would restrict the coded exposure or coded aperture to binary values, for instance because those would be simpler to realize physically, the proposed codification can still produce spatio-spectral patterns with non-binary values of attenuation. The bottom row of Fig.~\ref{fig:tmca_codes} depicts the codification of the proposed TMCA.
\subsection{Conditioning of the measurement matrices}
\label{subsec:conditioning}
We empirically analyze the conditioning of the proposed TMCA codifications. To do so, we consider discretized versions of Equations~\eqref{eq:CSI_prop_cont_4} and~\eqref{eq:LF_prop_cont_2} (derived in the supplemental). Both reduce to the general form presented in Equation~\eqref{eq:exposure_coded_matrix}, where $\bold{M}$ is the measurement matrix. The eigenvalue distribution of this matrix informs us about its conditioning and thus our ability to solve the inverse problem, that is recovering $\bold{x}$ from $\bold{e}$. 

Figure~\ref{fig:eigenvalues} shows the two TMCAs for different number of discrete slots $K$ in the shutter function (Section~\ref{subsec:tmca_generalities}). The cases shown for $K=1$ are equivalent to using no shutter function (a single slot means a single integration). Those are the measurement matrices induced by the traditional compressive light field in \cite{marwah2013compressive} and compressive spectral imaging in \cite{wagadarikar2008single}. The plots for both applications show a better conditioning of the measurement matrices for $K=8$ (TMCA) compared with $K=1$ (traditional CA). The ratio between the lowest and highest eigenvalues is lower for $K=8$ and the distribution is more uniform (the eigenvalues decay less rapidly), indicating TMCA is a better codification.

\section{End-to-end optimization: learning codes and reconstructions jointly}
\label{sec:implem}
We consider an end-to-end approach in which we optimize both the optical electronic encoder realizing the TMCA with a NN decoder solving the inverse problem (Figure~\ref{fig:teaser}). Using NNs to implement our differentiable decoders presents three main advantages: 1) they are differentiable: the error is propagated back to the codes and can jointly optimize both the encoder and decoder, 2) they embed priors that are task and dataset dependent, 3) they use a single feed-forward pass which is, in many cases, faster than traditional iterative methods (such as dictionary based methods).
\begin{figure}[t!]
	\centering
	\includegraphics[width=\linewidth]{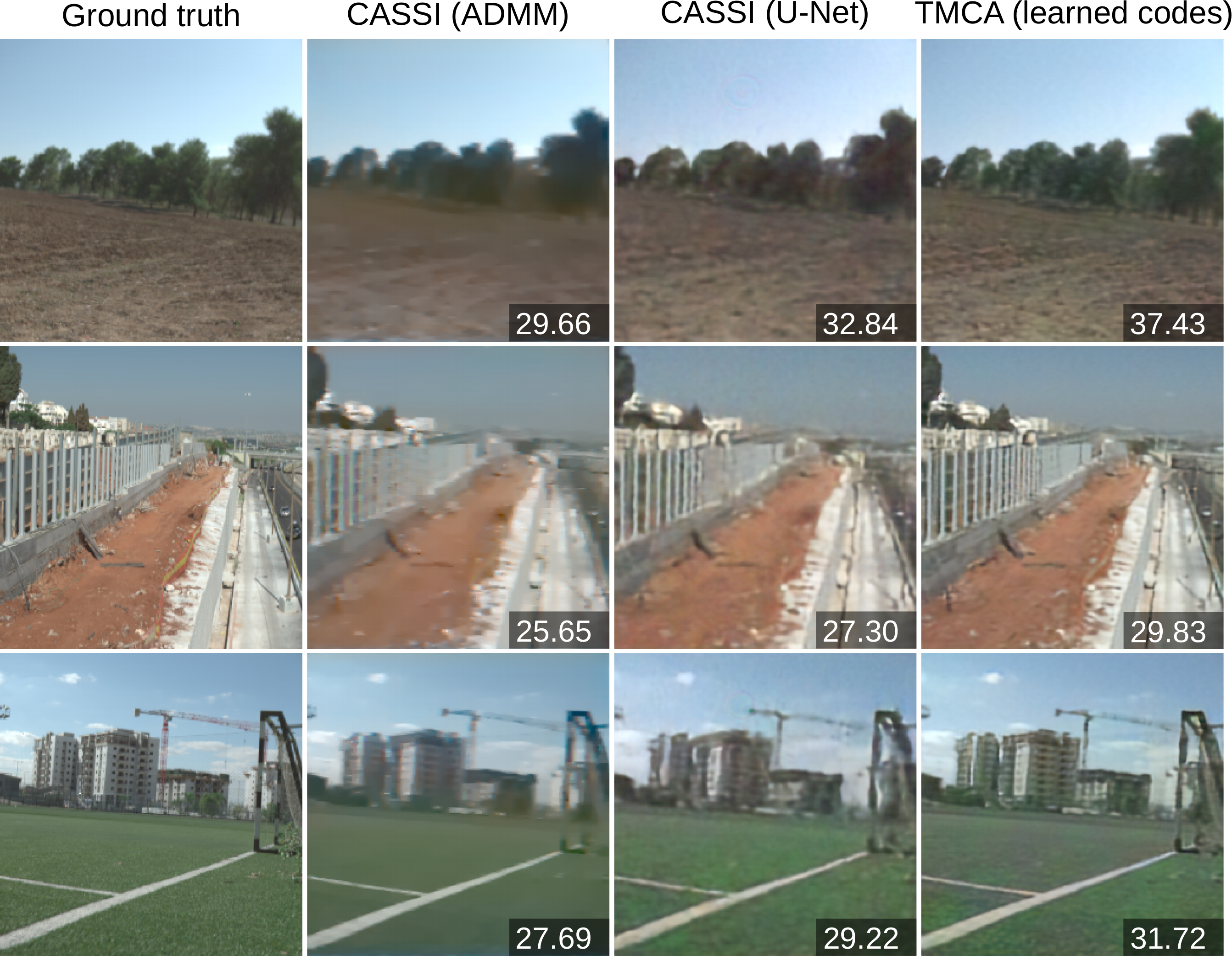}
	\caption{\small Examples of compressive spectral imaging reconstructions in simulation. The numbers in the lower-right corners are PSNR in dB comparing the reconstructions to ground truth images.\vspace{-1.5em}}
	\label{fig:spectral_results}
\end{figure}
\newcommand{\resiz}[1]{\footnotesize{#1}}
\begin{table}[t!]
	\centering
	{\small
	\setlength{\tabcolsep}{3pt}
	\begin{tabular}{cccccc}
		\toprule
		\makecell{Methods} & \resiz{PSNR}($\uparrow$)  & \resiz{UIQI}($\uparrow$)  & \resiz{SAM}($\downarrow$)   &\resiz{ERGAS}($\downarrow$) & \resiz{DD}($\downarrow$) \\
		\midrule
		\makecell{CASSI \resiz{(ADMM)}} & {27.40} & {0.938} & {22.42} & {9.56} & {0.031} \\
		\makecell{CASSI \resiz{(U-Net)}} & {29.66} & {0.968} & {15.99} & {7.04} & {0.022} \\	
		\makecell{TMCA \resiz{(random)}} & {31.39} & {0.978} & {13.01} & {5.74} & {0.019} \\
		\makecell{TMCA \resiz{(learned)}} & \textbf{32.72} & \textbf{0.981} & \textbf{11.92} & \textbf{5.27} & \textbf{0.016} \\			
		\bottomrule
	\end{tabular}%
	}
	\caption{Compressive spectral imaging: Comparison of the proposed TMCA with baselines on the ICVL 1 dataset.}
	\label{tab:CSI_ICVL}%
\end{table}%
%
\vspace{-.5em}
\paragraph*{Encoders} 
There are two main challenges to address in the end-to-end optimization of our TMCA. First, since both the CA and shutter functions are implemented in hardware (see Section~\ref{sec:results}), they are subject to the real devices' and systems' constraints. In particular our CAs and shutter functions must be binary valued. A second challenge is that we discretized the time domain in time slots to easily synchronize them. This essentially means the parameters representing those functions need to encode the slot at which sub-exposure start and stop. Without any further constraint on the functional form of the CA or coded exposure, the number of those parameters grows as the number of slots increases, preventing us from using too many slots.

To optimize the TMCA under those constraints we use an approach similar to the one presented in \cite{martel2020neural}. A forward pass through the encoder implements the exact discrete, binary valued model but the backward pass used to optimize its parameters is mismatched and implemented considering a continuous approximation of the discrete forward model. As an example, hard thresholds used for quantization in the forward pass are considered to be sigmoid functions in the backward pass. This ``forward-backward mismatch" approach enables gradients to flow backward to update the encoder's parameters while keeping an exact forward model that can be directly translated into hardware.

In the following, we denote our encoders as $\bold{M}_\phi$, where $\phi$ are the set of parameters that encode the discrete forward model. In TMCA, those parameters are essentially $K$ arrays of $M\times N$ pixels representing the time varying CA (on $K$ slots), and the $K$ arrays of $M'\times N'$ sensor's pixel representing the shutter functions.
\vspace{-.5em}
\begin{figure}[t!]
	\centering
	\includegraphics[width=1\linewidth]{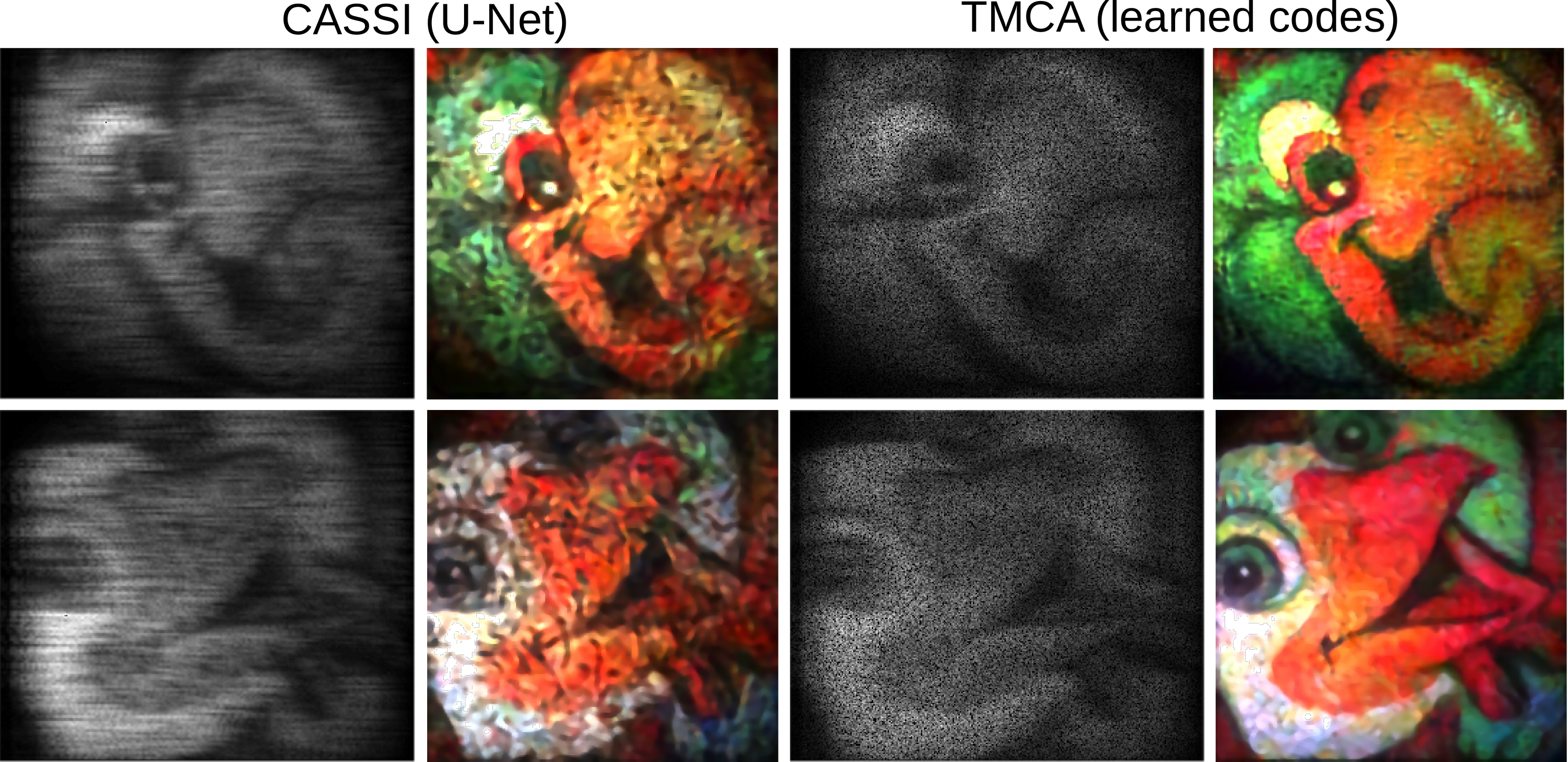}
	\caption{\small Examples of real captures of coded snapshots for hyperspectral imaging and their reconstructions.}
	\label{fig:coded_spectral_results}
\end{figure}
\paragraph*{Decoders} 
Many types of differentiable decoders have been considered for end-to-end optimization \cite{sitzmann2018end,wu2019phasecam3d,nehme2020deepstorm3d}. 
For the hyperspectral imaging application we use a vanilla U-Net preceded by a lifting of the measurement using a back-projection with the transposed of the measurement matrix. The measurement $\mathbf{e}$ generated by our hyperspectral encoder is a single 2D snapshot of size $M\times (N+L-1)$ while the hyperspectral image $\mathbf{x}$ we wish to reconstruct is a cube of size $M\times N \times L$. U-Nets operate by translating one domain to another of same dimension. Therefore, we first lift the measurements $\mathbf{e}$ in the spectral domain using the transpose operator $\mathbf{M}^T$ creating $\mathbf{e'}=\bold{M}^T\bold{e}$, which is then fed to the U-Net decoder to reconstruct the hyperspectral image $\mathbf{\tilde{x}}$. For the compressive light field application we used the unrolled optimization network proposed by \cite{guo2020deep}. We denote the function produced by those NNs $\mathcal{D}_{\psi}$ where $\psi$ are their learnable parameters.

\emph{At training time}, a ground-truth measurement $\mathbf{x}_{\text{GT}}$ is sampled from a dataset of $N$ hyperspectral images or light fields. It is then encoded via the forward model as $\bold{e}=\bold{M}_\varphi\bold{x}_{\text{GT}}$. The decoder proposes a reconstruction $\tilde{\bold{x}}=\mathcal{D}_\psi(\bold{e})$. To learn the parameters $\{\varphi,\psi\}$ of our encoder-decoder architecture we optimize the loss function $\mathcal{L}$ to minimize the discrepancy between the reconstructed signals $\tilde{\mathbf{x}}=\mathcal{D}_\psi \circ \mathbf{M}_\varphi ( \mathbf{x}_{\text{GT}}^n)$ and the ground-truth $\mathbf{x}_{\text{GT}}$:
\begin{align}
    \underset{\phi,\psi}{\text{argmin}} \sum_{n=1}^N \mathcal{L}\big( \mathbf{{x}}^n_{\text{GT}} , \mathcal{D}_\psi \circ \mathbf{M}_\varphi ( \mathbf{x}_{\text{GT}}^n )\big).
\end{align}
where $\circ$ stands for the composition of functions. In practice, we choose $\mathcal{L}$ to be a $L_2$ or $L_1$ norm, but this could also be chosen to be a high-level perceptual loss function such as a VGG loss \cite{simonyan2014very} or analogous.

\emph{At inference} in simulation, the pipeline is run with $\bold{x}_{\text{GT}}$ sampled from a test set. For real captures, the learnt parameters of the encoder are directly translated from simulation to hardware. This is possible because the encoders are, by construction, designed to emulate physically realizable CAs and shutter functions. 

\section{Results}
\label{sec:results}
\begin{figure}[t!]
	\centering
	\includegraphics[width=1\linewidth]{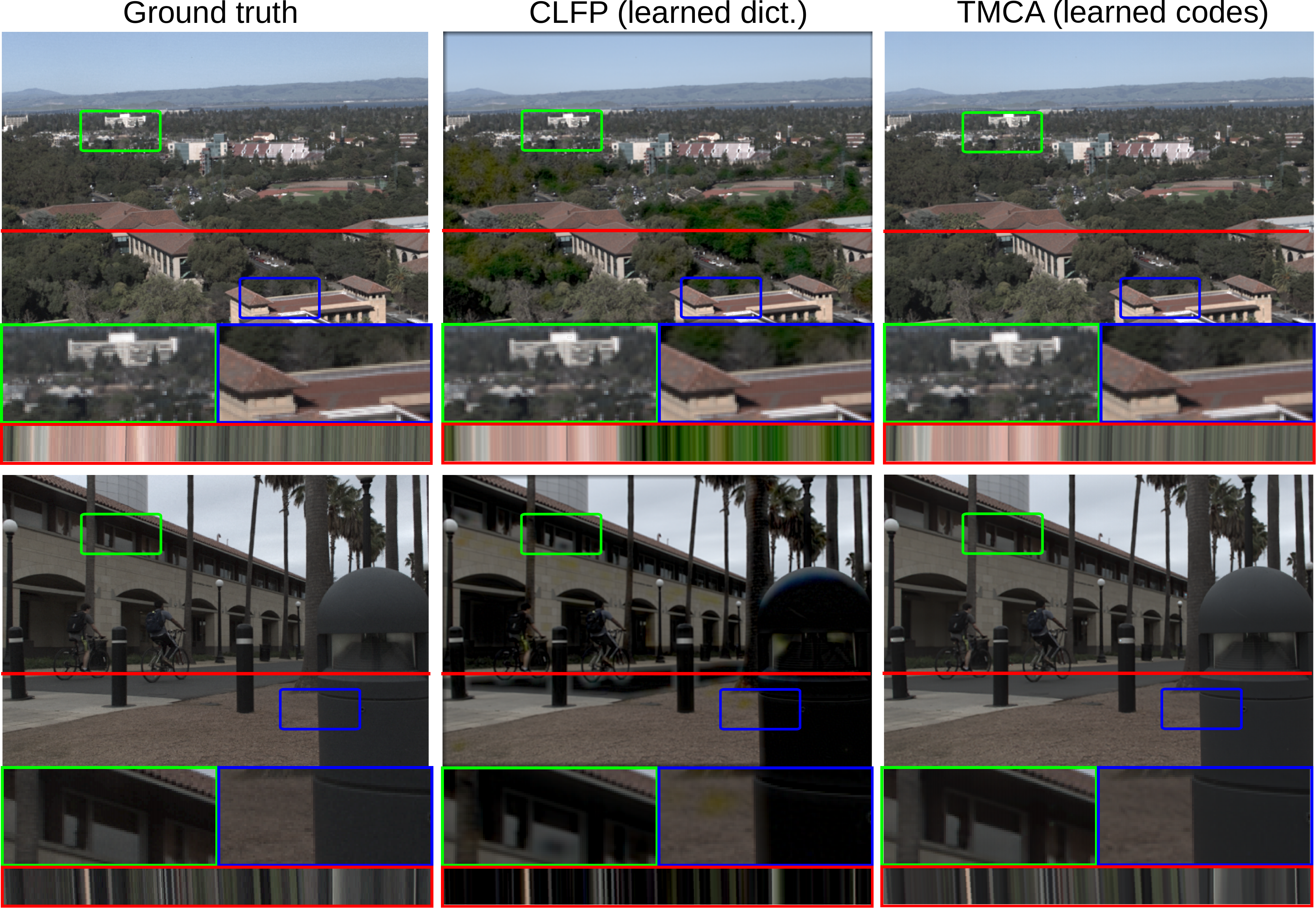}
	\caption{\small Examples of light field reconstructions (central view) in simulation (shown with zoom-ins) comparing CLFP \cite{marwah2013compressive} and our proposed TMCA with ground truth.\vspace{-1.0em}}
	\label{fig:coded_LF_syn_results}
\end{figure}
We show results of the proposed TMCA for the two applications we consider in this work: compressive light field and hyperspectral imaging. We demonstrate both results in simulations and on real captures taken with two prototype systems we built.
All our models are implemented in Pytorch \cite{paszke2017automatic} and trained on Titan X GPU using the ADAM optimizer \cite{kingma2014adam}, complete training details and photographs of our setups are presented in the supplemental.
\subsection{Coded hyperspectral Imaging}
\paragraph*{Simulations}
We learn the end-to-end model for hyperspectral imaging using the ICVL dataset. It consists of 200 spectral images. We randomly select 160 hyperspectral images for training, 20 for validation, and 20 for testing, cropped at a size of $256\times256$ with $L=12$ spectral bands. We set the number of time slots in the TMCA encoder to $K=8$. The U-Net is trained for 500 epochs.

We compare the proposed TMCA codification against three different baselines: a) the traditional CASSI codification using random binary patterns and reconstructed using the alternative direction method of multipliers (ADMM) b) the CASSI codification using a trained U-Net as a decoder c) the proposed TMCA codification and reconstruction pipeline using random (non-optimized) codes and d) our full TMCA codification with learnt codes. The quantitative results performed on our random test fold of ICVL are presented in Table~\ref{tab:CSI_ICVL}, they show that on all the metrics we evaluated on (PSNR, UIQI, SAM, ERGAS, and DD, see supplemental for details) our full TMCA pipeline performs better than all the other baselines.
%
We show qualitative results of a few reconstructed hyperspectral images in Figure~\ref{fig:spectral_results}. Those illustrate how the proposed TMCA appears closer to the ground truth both in terms of spatial accuracy (they are less blurry) as well as spectral accuracy (colors match better). 
%
%
%
\begin{table}[t!]
	\centering
	{\small
	\begin{tabular}{lrr}
		\toprule
		   Methods & PSNR($\uparrow$)   & SSIM($\uparrow$)   \\
		\midrule
		  {CLFP\cite{marwah2013compressive} (\resiz{learned dict.}})  & {30.06} & {0.82} \\
		  {CLFP\cite{marwah2013compressive} (\resiz{deep net.}\cite{guo2020deep}}) & {32.43} & {0.91}  \\	
		  {TMCA (\resiz{random codes}}) & {34.03} & {0.93}  \\	
		  {TMCA (\resiz{opt. codes}}) & \textbf{34.89} & \textbf{0.94}  \\			
		\bottomrule
	\end{tabular}%
	}
	\caption{Compressive light field imaging: comparison of the proposed TMCA against baselines using the aggregated Lytro dataset.
	}
	\label{tab:CLFI}%
\end{table}%
%
\paragraph{Real captures} We built a prototype of our system to evaluate the proposed TMCA approach for real-world scenes. The system consists of an achromatic objective lens with $50$mm focal length (Thorlabs AC254-050-A-ML), a digital micromirror device (DMD DLi4120), an F/8 relay lens, a custom double Amici prism with center wavelength $550$nm captured through a monochrome CCD sensor Stingray F-080B with $4.65\mu$m pixel size. The same prototype was used to capture measurements for the traditional CASSI codification and the proposed TMCA. The spectral images are reconstructed using the U-Nets trained in simulation, but using the real measurements. Results showing the capture of a book cover are shown in Figure~\ref{fig:coded_spectral_results} demonstrating that TMCA imaging can be implemented in a real-world hyperspectral imaging prototype that still presents a higher visual quality than the traditional CASSI.
\subsection{Compressive Light Field Imaging}
\paragraph{Simulations}
We learn our end-to-end TMCA for compressive light field imaging on a dataset aggregating real-world and synthetic light fields (LF). We use 100 real captured LF of size $7\times 7 \times 376 \times 541$ from the Kalantari et al. Lytro dataset \cite{kalantari2016learning}, 22 synthetic LF images of size $5\times5\times512\times512$ from \cite{shi2019framework}, and 33 synthetic LF images of size $5 \times 5 \times 512 \times 512$ from \cite{honauer2016dataset}. 
We randomly split this aggregated dataset in 110 LF images for training, 20 for validation, and 25 for testing. 

In our experiment, we reconstruct $5\times 5$ angular views from a single snapshot with a resolution of $480 \times 270$ pixels. We use randomly cropped patches of those images of spatial size $11 \times 11$ and consider $5 \times 5$ angular views for training: randomly cropping the 4D LF into patches increases the number of samples at training while reducing the memory requirements to process an entire light field. The decoder is the deep spatial-angular convolutional sub-network proposed in \cite{guo2020deep}, it is trained for 500 epochs. After training, we reconstruct overlapping 4D patches that are then merged with a median filter.

We compare the proposed TMCA codification against three different baselines similar to the compressive imaging application: a) a reconstruction using the traditional dictionary learning and reconstruction approach of \cite{marwah2013compressive}, b) the same codification but using the deep network decoder from \cite{guo2020deep} c) our TMCA with random (non-optimized codes) using the deepnet decoder of \cite{guo2020deep} d) our full TMCA codification pipeline with learnt codes. The results are compiled in Table~\ref{tab:CLFI}, showing that on the two metrics we evaluated on (PSNR and SSIM) the TMCA is superior to the other baselines. Qualitative results are shown in Figure \ref{fig:coded_LF_syn_results} where two recovered LFs of the testing set (Kalantari Lytro LFs) are reconstructed and compared with the ground truth. 
%
\paragraph*{Real captures}
We assess the proposed TMCA approach in a real experiment. We use a liquid crystal on silicon (LCoS) display (HOLOEYE PLUTO-2.1 LCoS SLM) where each pixel can independently change the polarization state of the incoming light field, in conjunction with a polarizing beam splitter and relay optics. As a single pixel on the LCoS cannot be well resolved with our setup, we treat blocks of $4\times4$ LCoS pixels as macro pixels, resulting in a CA of $480\times 270$. We reimage the LCoS with an SLR camera lens (Canon EF-S $18-55$ f/$4-5.6$ IS STM) which is not focused on the LCoS but in front of it, thereby optically placing the (virtual) image sensor behind the LCoS plane. A Canon EF $80$ mm, f/$5.6$ II lens is used as the imaging lens and focused at a distance of $60$cm. We follow the same procedure as in \cite{marwah2013compressive} to adjust the distance between the mask (LCoS plane) and the virtual image sensor for capturing light fields with $5\times 5$ angular resolution. 

\begin{figure}[t!]
	\centering
	\includegraphics[width=0.99\linewidth]{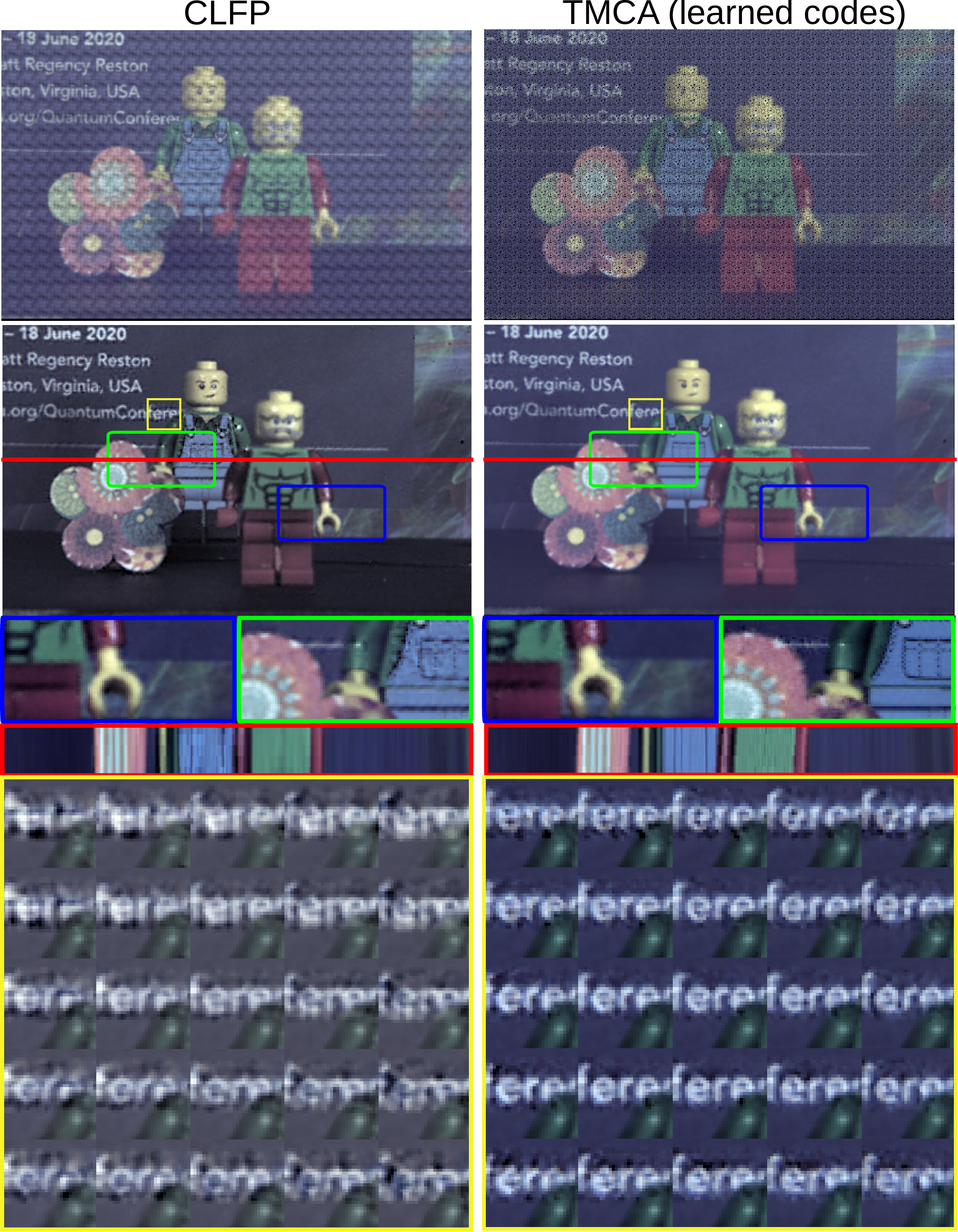}
	\caption{\small Examples of real captures of coded snapshots for light field comparing CLFP \cite{marwah2013compressive} and of our proposed TMCA and their reconstructions.}
	\label{fig:results_LF_real}
\end{figure}

The real captured measurements are reconstructed using the same deep networks as in simulation. The measurements and reconstructions are shown in Figure~\ref{fig:results_LF_real} which shows that the TMCA enables reconstructions featuring better spatial resolution in the angular views.
%

\section{Discussion}
\label{sec:discussion}
We introduced a new coding strategy dubbed Time Multiplexed Coded Aperture (TMCA) and demonstrated it improves CA codifications in two applications: compressive hyperspectral imaging and light field imaging. TMCA improves upon traditional CA without introducing additional optical elements. It adds a coded shutter synchronized with the CA. The coded shutter can be simply realized electronically, using dedicated sensors or burst of multiple snapshots averaged together. We optimized TMCA codes in an end-to-end optimization approach and demonstrated in both synthetic and real experiments that the proposed TMCA yields largely superior reconstruction quality compared with traditional CA approaches. While a limitation of CA systems are their low light efficiency, which could be detrimental in low-light scenarios or high-speed imaging, we believe our system could be further engineered to be real-time (our captures for $K=8$ take $8\times20$ms and reconstructions take about a second), thus enabling high-quality videos in those applications. We believe the proposed TMCA would naturally extend to many other applications that make use of CAs and improve their reconstruction quality.   

\section*{Acknowledgments}
J.N.P.M. was supported by a Swiss National Foundation (SNF) Fellowship (P2EZP2\_181817), G.W. was supported by an NSF CAREER Award (IIS 1553333), and a PECASE by the ARL. H.A. was supported by the Fullbright 2019 Visiting Scholar Program.

{\small
\bibliographystyle{ieee_fullname}
\bibliography{bibliography}
}

\clearpage

\pagebreak
\setcounter{equation}{0}
\setcounter{figure}{0}
\setcounter{table}{0}
\setcounter{section}{0}
\setcounter{page}{1}
\makeatletter
\renewcommand{\theequation}{S\arabic{equation}}
\renewcommand{\thefigure}{S\arabic{figure}}

%
%
%
%
%

\title{Time-Multiplexed Coded Aperture Imaging: Learned Coded Aperture and Pixel Exposures for Compressive Imaging Systems\\
--Supplemental Material--}

\author{
Edwin Vargas$^{1,*}$, Julien N.P. Martel$^{2,*}$, Gordon Wetzstein$^2$, Henry Arguello$^1$\\
$^1$Universidad Industrial de Santander, Colombia $^2$Stanford University, USA\\
{\tt\small edwin.vargas4@correo.uis.edu.co, jnmartel@stanford.edu} \\
{\tt\small gordon.wetzstein@stanford.edu, henarfu@correo.uis.edu.co}
}

\newcommand{\thefootnote}[1]{}
\maketitle
\ificcvfinal\thispagestyle{empty}\fi

\newcommand{\eqclf}{(\textcolor{red}{8}) }
\newcommand{\eqcsi}{(\textcolor{red}{11}) }
\newcommand{\eqcsitwo}{(\textcolor{red}{12}) }
\newcommand{\normtwos}[1]{\lVert #1 \rVert_2}

\section{Derivation of the discrete forward models}
In this section, we expose the discrete forward models from the continuous models shown in Equation~\eqclf and~\eqcsi of the main paper. 
\subsection{Compressive Light Field Imaging}
We recall the continuous forward model for compressive light field imaging from Equation~\eqclf: 
\begin{equation}
\label{eq:LF_prop_cont}
e(x) = \int_{\mathcal{V}} \int_{\Delta{t}} S(x,t')\,l(x,u)\,T(x + s(u-x),t') \dd{u} \dd{t'}.\nonumber
\end{equation}
Considering $K$ discrete time slots, the discretized form of the coded exposure for a given pixel $m$ in the sensor array, can be written as:
\begin{align}
e_{m} =  \sum_{k=1}^K\sum_{\ell=(U-1)/2}^{(U+1)/2} S^k_{m}\,T^k_{m+\ell}\,l_{m,\ell},
\label{eq:discrete_CLF}
\end{align}
where $m,\ell$ and $K$ are the indexes for the discretized spatial, angular and time dimensions, and $M,U$ and $K$ are the corresponding number of samples along these dimensions. Note the coded aperture is defined as shifted by a given number of pixels depending on the sub aperture image~$\ell$. 

We define the discrete TMCA as 
\begin{align}
    \hat{T}_m = \sum_{k=1}^K S^k_{m}\, T^k_{m+\ell},
\end{align}
and the discrete model of the coded exposure in \eqref{eq:discrete_CLF} can be finally expressed as
\begin{align}
    e_{m} =  \sum_{k=1}^K\sum_{\ell=(U-1)/2}^{(U+1)/2} \hat{T}_{m,\ell}\, l_{m,\ell}.    
\end{align}
%
\subsection{Compressive spectral imaging}
The continuous forward model for the coded exposure of our compressive spectral imaging system is given in Equation~\eqcsi of the main paper:
\begin{align}
e(x,y) = \int_{\Delta t} S(x,y,t')\iiint T(x',y',\lambda,t')\,I(x',y',\lambda)\nonumber\\ 
h(x-\mathcal{S}(\lambda) - x',y-y')\,\kappa(\lambda) \dd{x'}\dd{y'}\dd{\lambda}\dd{t'}.\nonumber
\end{align}
\begin{figure}[t!]
    \centering
    \includegraphics[width=\linewidth]{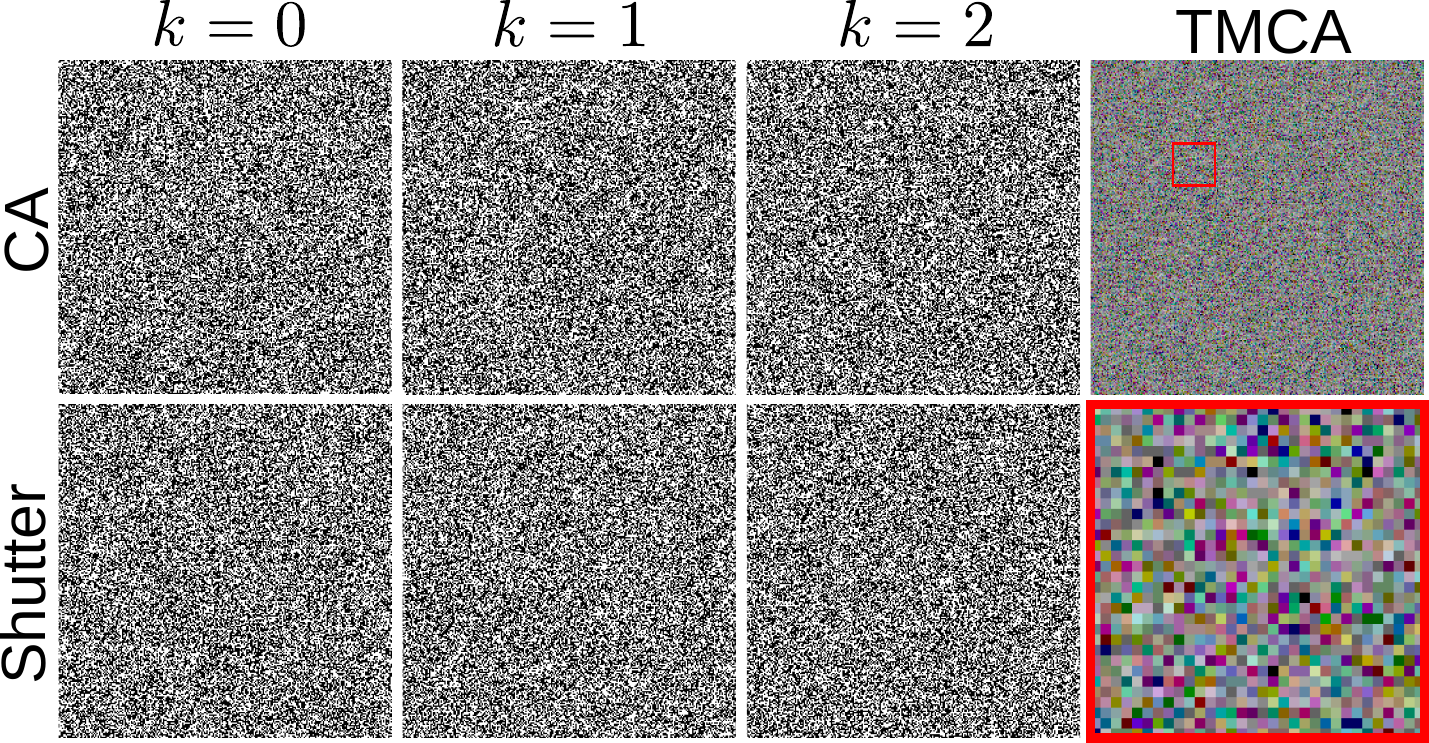}
    \caption{Learned TMCA for compressive spectral imaging. Learned CA and shutter function for the three first time slots $k=0,1,2$. The resultant learned TMCA in CSI is equivalent to a colored coded aperture shown in the last column of the figure.}
    \label{fig:coded_learned_spectral}
\end{figure}

For a given pixel $(m,n)$ (remember we now index the spatial location in 2D since one of the spatial dimension is ``special" and corresponds to the dimension in which the prism disperses light), and considering $K$ discrete time slots, the measurement model can be written as:
\begin{align}
e_{m,n} =  \sum_{k=1}^K\sum_{\ell=1}^{L} \sum_{i=0}^{M-1} \sum_{j=0}^{N-1} S^k_{m,n}  \,f_{i,j,\ell}\, T^k_{i,j}\, h_{m-i,n-j,\ell} 
\label{eq:discrete_CASSI2}
\end{align}
where $i,j$ are indices along the discretized spatial dimensions of $M,N$ samples each, $\ell,k$ denote the indices for the discretized wavelength and time dimensions, and $L,K$ are the corresponding number of samples along those.

The point spread function (PSF) $h$ corresponds to a propagation model through a unit magnification imaging optics. Further assuming the prism features linear dispersion, the PSF can be expressed as the shifted dirac $\delta_{m-i,n-j,\ell}$. Substituting this expression in Equation~\eqref{eq:discrete_CASSI2}, and simplifying, the exposure model becomes 
\begin{align}
e_{m,n} =  \sum_{k=1}^K\sum_{\ell=1}^{L} \sum_{i=0}^{M-1} \sum_{j=0}^{N-1} S^k_{i,j+\ell} \,T^k_{i,j}\, f_{i,j,\ell}\, \delta_{m-i,n-j,\ell},
\label{eq:discrete_CASSI3}
\end{align}
Grouping the time variables in \eqref{eq:discrete_CASSI3}, we then define the discrete TMCA:
\begin{align}
\hat{T}_{i,j,\ell} =\sum_{k=1}^K S^k_{i,j+\ell}\, T^k_{i,j}.
\label{eq:TMCA_S}
\end{align}
yielding the coded measurements
\begin{align}
e_{m,n} = \sum_{\ell=1}^{L} \sum_{i=0}^{M-1}\sum_{j=0}^{N-1} \hat{T}_{i,j,\ell}\, f_{i,j,\ell}\, \delta_{m-i,n-j,\ell} 
\label{eq:discrete_CASSI4}
\end{align}
%
\section{Deriving Equation~\eqcsitwo from Equation~\eqcsi}
Again, Equation~\eqcsi is:
\begin{align}
e(x,y) = \int_{\Delta t} S(x,y,t')\iiint T(x',y',t')\,I(x',y',\lambda)\nonumber\\ 
h(x-\mathcal{S}(\lambda) - x',y-y')\,\kappa(\lambda) \dd{x'}\dd{y'}\dd{\lambda}\dd{t'}.
\nonumber
\end{align}
Since $h$ is the propagation through unit magnification imaging optics and a dispersive element with linear dispersion, the impulse response can be expressed as $h(x-\mathcal{S}(\lambda) - x',y-y')=\delta(x-\lambda - x',y-y')$, and the coded exposure can be simplified as
\begin{align}
e(x,y) = \int_{\Delta t} \int S(x,y,t')\,T(x-\lambda,y,t')\,\nonumber\\ 
I(x-\lambda,y,\lambda)\,\kappa(\lambda)\dd{\lambda}\dd{t'}.
\end{align}
Using the properties of dirac distributions, we express the terms inside the integral as the following convolution 
\begin{multline}
S(x,y,t')\,T(x-\lambda,y,t') I(x-\lambda,y,\lambda)= \iint S(x'+\lambda,y',t')\,\\
T(x',y',t')\,I(x',y',\lambda)\,\delta(x-\lambda - x',y-y')\,\dd{x'}\dd{y'}.
\end{multline}
Substituting this expression in Equation~\eqcsi, we obtain the following expression (Equation~\eqcsitwo in the main paper) for coded exposure measurements:
\begin{align}
\label{eq:CSI_prop_cont_3}
e(x,y) = \int_{\Delta t} \iiint S(x'+\lambda,y',t')\,T(x',y',t')\,I(x',y',\lambda)\nonumber\\ 
\delta(x-\lambda - x',y-y')\,\kappa(\lambda) \dd{x'}\dd{y'}\dd{\lambda}\dd{t'}.
\end{align}
\begin{figure}[t!]
    \centering
    \includegraphics[width=\linewidth]{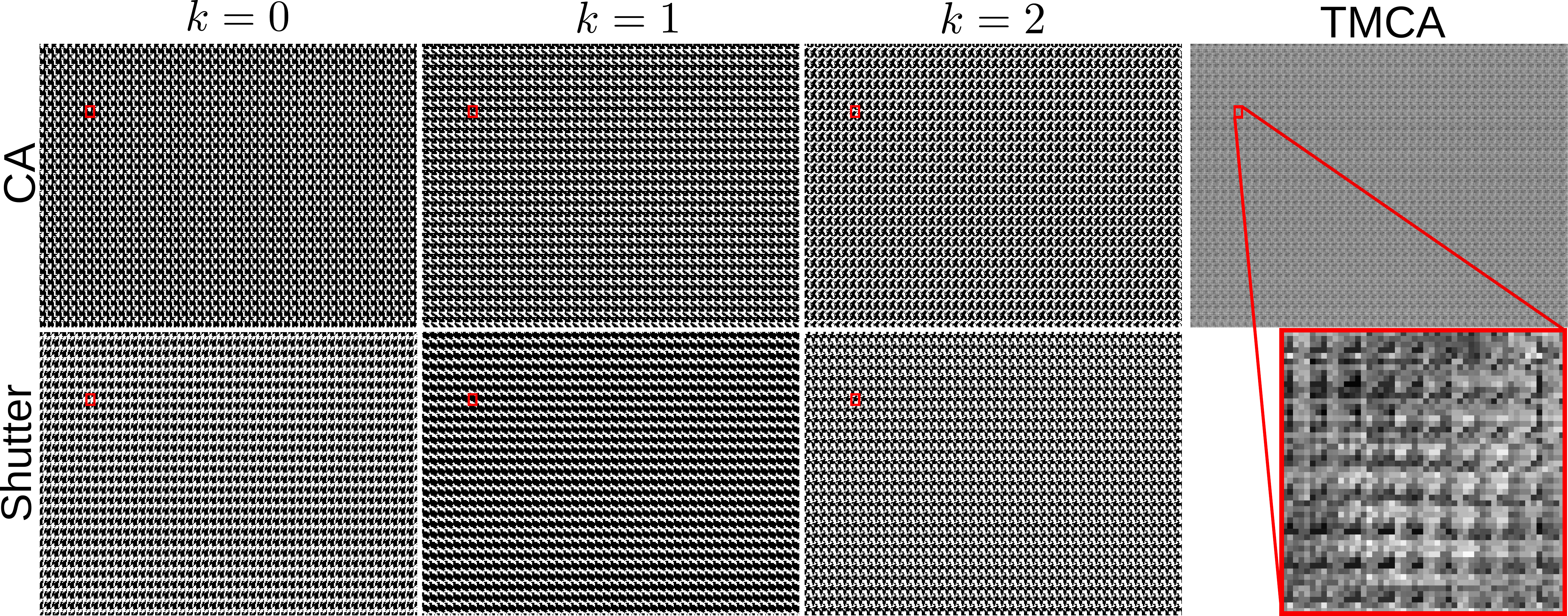}
    \caption{Learned TMCA for compressive light field imaging. Learned CA and shutter function for the first three time slots $k=0,1,2$. We use a 2D array of sub-images for visualization of the TMCA where each sub-image represents the response of the equivalent coded aperture to all rays arriving at one point on the coded aperture from all points on the aperture plane. Thus, the resultant learned TMCA is equivalent to a coded aperture with sensitive angular pixels. }
    \label{fig:coded_learned_lf}
\end{figure}
\begin{figure*}[t!]
    \centering
    \includegraphics[width=\linewidth]{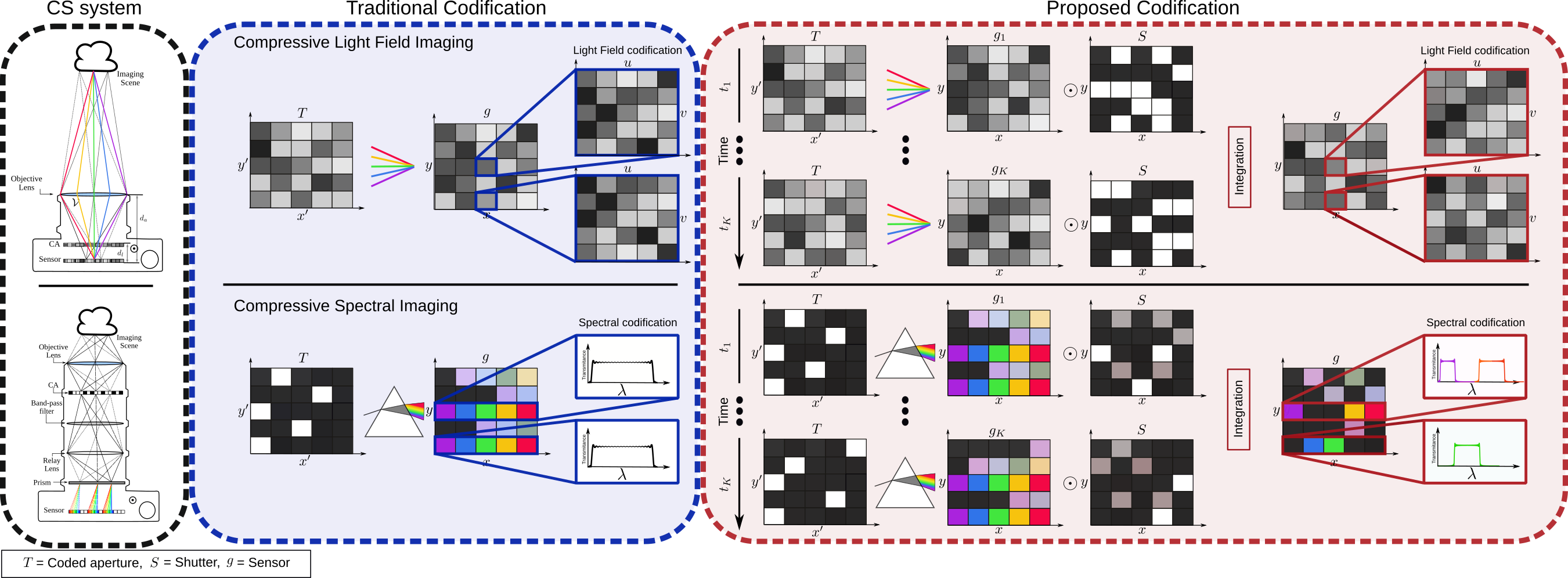}
    \caption{A diagram representing the advantages of the proposed TMCA codification against traditional CA codifications for the two applications we study: compressive light field imaging and compressive hyperspectral imaging.}
    \label{fig:diagram_codification}
\end{figure*}
%
\section{Metrics used for spectral imaging}
Here we give the formal definitions and some intuition about the metrics we used to evaluate the quality of our compressive spectral imaging system.
\begin{itemize}\itemsep0em 
	\item \textit{RMSE:} The root mean square error (RMSE) is a pixel-wise dissimilarity measure between a ground-truth spectral image $\mathbf{x}$ and the reconstructed image $\mathbf{\hat{x}}$ of $N\times M$ pixels and $L$ spectral bands defined as	
	\begin{equation}
	    	\text{RMSE}(\mathbf{x},\mathbf{\hat{x}}) = \frac{1}{M\cdot N\cdot L} ||\mathbf{{x}}-\mathbf{\hat{x}}||_2. 
	\end{equation}

	\item \textit{UIQI:} The universal image quality index (UIQI) was proposed in \cite{wang2002universal} for evaluating the similarity between two single gray scale images. This metric measure the correlation, contrast and luminance distortion of a reconstructed image with respect to the reference image. The UIQI between two single-band images $\mathbf{a}=[a_1,\cdots, a_{MN}]$ and $\hat{\mathbf{a}}=[\hat{a}_1,\cdots, \hat{a}_{MN}]$ is defined
	\begin{equation}
	\text{UIQI}({\mathbf{a}},\hat{\mathbf{a}})=\frac{4\sigma_{a\hat{a}^2}\mu_a\mu_{\hat{a}}}{(\sigma_a^2 + \sigma_{\hat{a}}^2)(\mu_a^2 + \mu_{\hat{a}}^2)}
	\end{equation} 
	where $(\mu_a, \mu_{\hat{a}}, \sigma_a^2, \sigma_{\hat{a}}^2)$ are the sample means and variances of $\mathbf{a}$ and $\hat{\mathbf{a}}$, and $\sigma_{a\hat{a}^2}$ is the sample covariance of $(\mathbf{a}, \mathbf{\hat{a}})$. The range of UIQI is $[-1,1]$ and $\text{UIQI}({\mathbf{a}},\hat{\mathbf{a}})=1$ when $\mathbf{a}=\hat{\mathbf{a}}$. Thus, the higher the UIQI, the better spectral reconstruction. Since we work with multi-band images, the overall UIQI metric reported on Table~I in the main paper corresponds to the average of the UIQIs over all spectral bands.
	
	\item \textit{SAM:}  The spectral angle mapper (SAM) was proposed to evaluate the quality of the recovered spectral images by measuring the similarity between reference and estimated spectral signatures \cite{kruse1993spectral}. The SAM of two spectral vectors $\mathbf{x}$ and $\hat{\mathbf{x}}$ is defined as
	\begin{equation}
	    \text{SAM}({\mathbf{x}},\hat{\mathbf{x}}) = \text{arccos}\left(\frac{\langle{\mathbf{x}},\hat{\mathbf{x}}\rangle}{\normtwos{\mathbf{x}}\normtwos{\hat{\mathbf{x}}}}\right).
	\end{equation}
 The SAM metric reported in Table I is obtained by averaging the SAMs computed from all $M\cdot N$ image pixels. Since the SAM is an angular quantity, the value of SAM is expressed in degrees and thus belongs to $(-90, 90]$. The smaller the absolute value of SAM, the more higher the spectral similarity is between the recovered image and the ground truth.
	
	\item \textit{ERGAS:} The relative dimensionless global error in synthesis (ERGAS) has been proposed to compute the amount of spectral distortion in super resolved spectral images \cite{wald2000quality}. Here, we employ this quantity to evaluate the recovered spectral images:
	\begin{equation}
	\text{ERGAS}=100\sqrt{\frac{1}{L}\sum_{i=0}^{L-1} \left( \frac{\text{RMSE}(\mathbf{x}_i,\hat{\mathbf{x}_i})}{\mu_i}\right)^2}
	\end{equation} 
	
	where $\mu_i$ is the mean of the $i$-th band ($\mathbf{x}_i$) of the spectral image, and $L$ is the number of spectral bands. The smaller ERGAS, the smaller the spectral distortion.
	
	\item \textit{DD:} The final metric reported in Table I is the degree of distortion (DD) between two spectral images which is defined as
	\begin{equation}
	    \text{DD}({\mathbf{x}},\hat{\mathbf{x}})= \frac{1}{N\cdot M\cdot L} \lVert \mathbf{x}-\hat{\mathbf{x}}\rVert_1.
	\end{equation}
	The smaller DD, the better the recovered spectral image.
\end{itemize}
\begin{figure*}[t!]
    \centering
    \includegraphics[width=\linewidth]{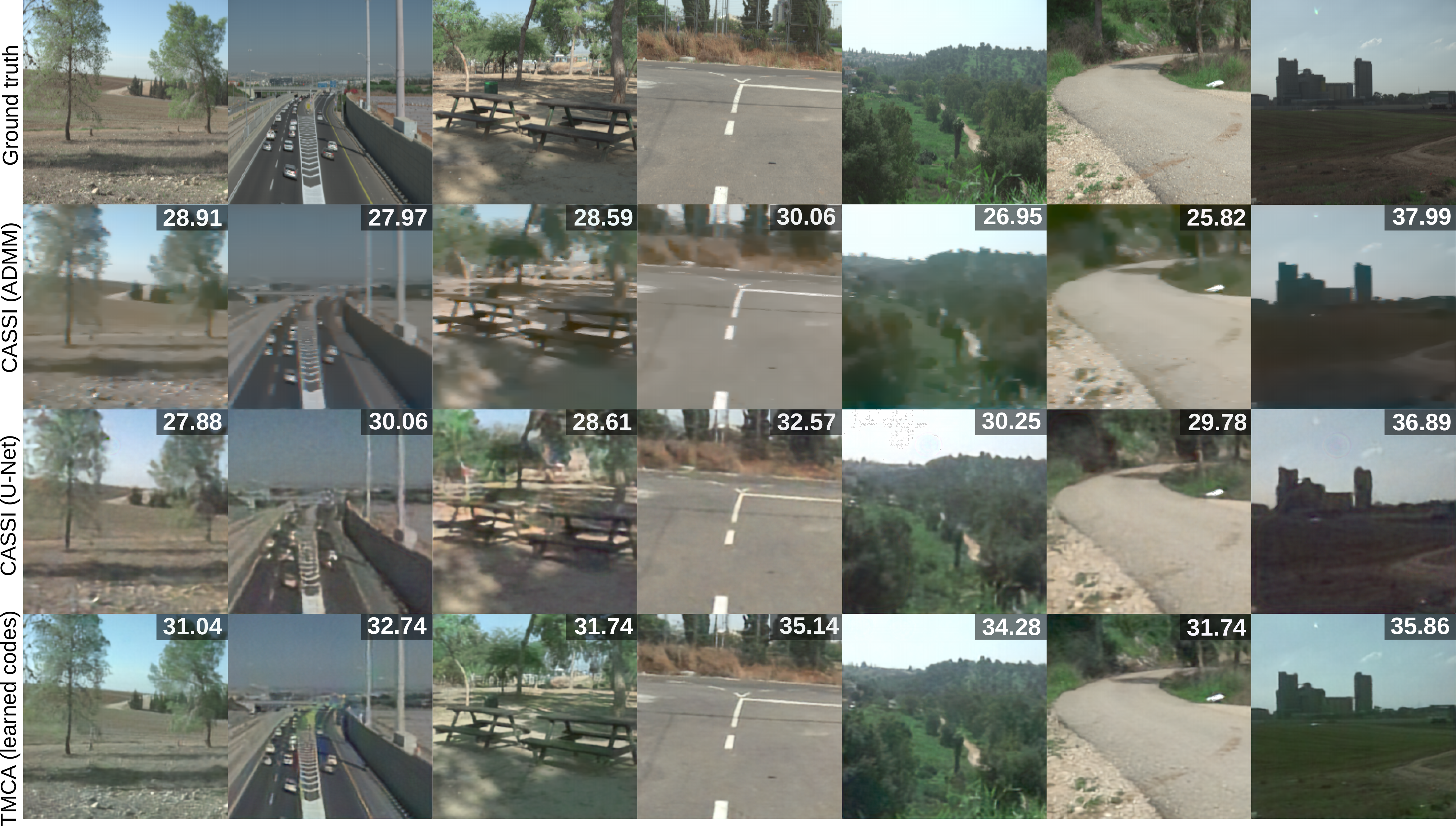}
    \caption{Additional results for compressive spectral imaging on the ICVL 1 test fold comparing our TMCA (fourth row) against the CASSI and using a conventional optimization technique (ADMM) as a decoder (second row) and a U-Net decoder (third row).}
    \label{fig:spectral_results}
\end{figure*}
%
\section{Simulation details}
In this section we present additional qualitative results of the simulations in compressive hyperspectral imaging and compressive light field imaging.  
\subsection{Coded Hyperspectral Imaging}
\paragraph*{Training details:} We learn the end-to-end model for hyperspectral imaging using 160 spectral images from the ICVL dataset \cite{arad_and_ben_shahar_2016_ECCV}. We used cropped images at a size of $256\times256$ with $L=12$ spectral bands. The $L$ spectral bands corresponds to the following wavelengths in nm: $[480, 500, 510, 530, 550, 560, 570, 590, 600, 620, 640, 650]$. We set the number of time slots in the TMCA encoder to $K=8$. The U-Net is trained for 500 epochs using ADAM optimizer. We applied a learning rate decay of factor $0.5$ every $150$ epoch with an initial rate of $0.0001$. We display the coded apertures and shutter function learned in our pipeline in Figure~\ref{fig:coded_learned_spectral}.

\paragraph{Rational behind our baselines,} we compare the proposed TMCA codification against three different baselines: a) the traditional CASSI codification using random binary patterns and reconstructed using the alternative direction method of multipliers (ADMM) b) the CASSI codification using a trained U-Net as a decoder c) the proposed TMCA codification and reconstruction pipeline using random (non-optimized) codes and d) our full TMCA codification with learned codes.

Baseline a) shows the performance of a system that does not use our codification, and uses a conventional optimization technique as a decoder. The baseline b) still uses the traditional codification but now uses a modern NN decoder, thus allowing us to state that the results we observe with baseline c) that uses our TMCA codification cannot only be attributed to the fact we use a NN decoder, but indeed that the TMCA codification is better. Finally, d) shows that optimizing the TMCA codification itself is also beneficial.

\paragraph{Additional qualitative results} We show qualitative results for a few more reconstructed hyperspectral images in Figure~\ref{fig:spectral_results}.

\paragraph{Mapping spectral bands to RGB } Our hyperspectral images are mapped to an RGB composite image by selecting three spectral channels corresponding to wavelengths of $650$,$550$ and $480$ nm.

\begin{figure}[t!]
    \centering
    \includegraphics[width=\linewidth]{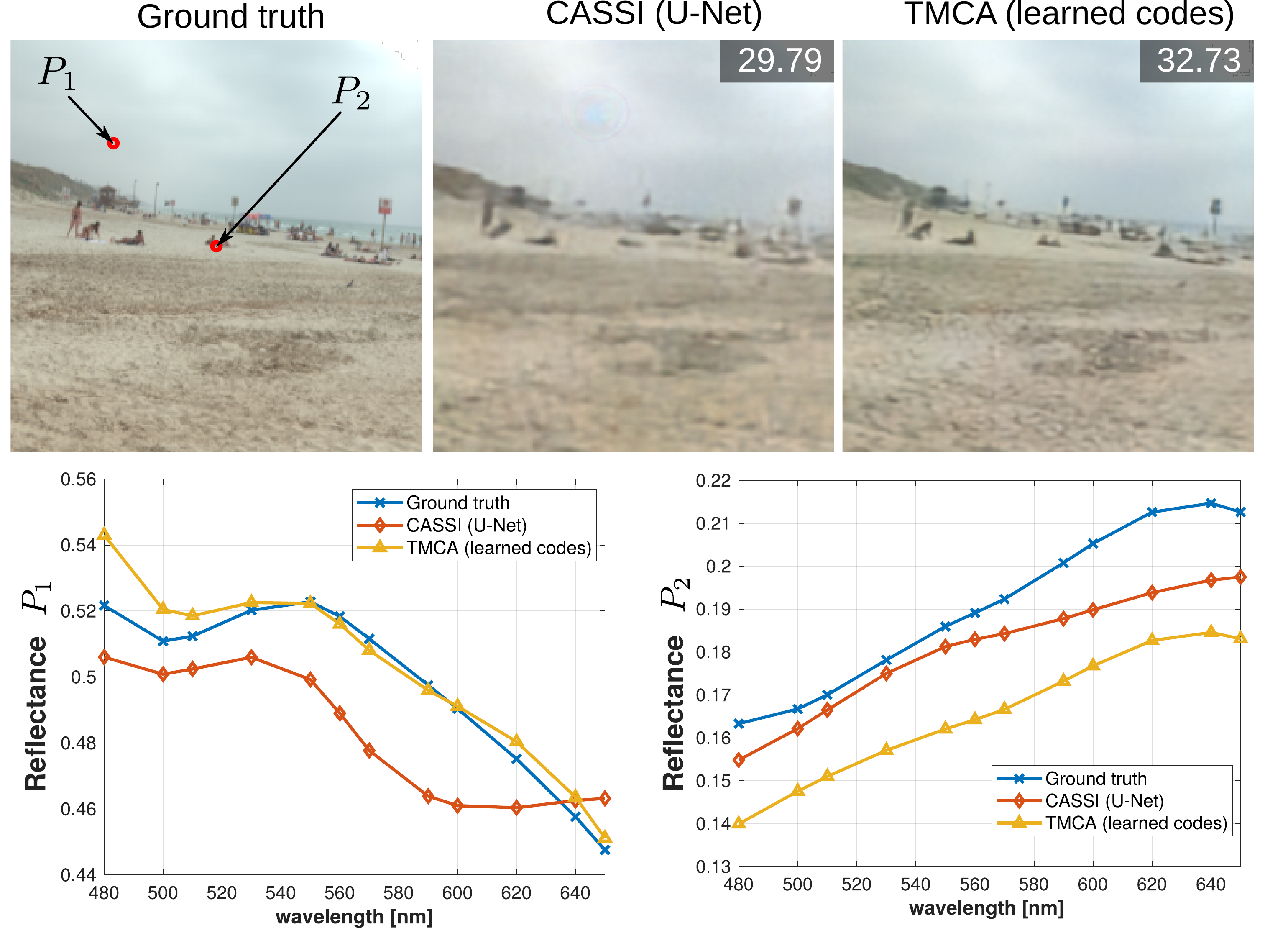}
    \caption{Spectral signatures for two points in an image randomly sampled from the ICVL 1 dataset. The $L=12$ spectral bands are shown for a point in the sky and a point in the sand on the reconstructions performed by our method using TMCA and learned codes and the conventional CASSI with a U-Net.}
    \label{fig:spectral_signatures}
\end{figure}

\begin{figure*}[t!]
    \centering
    \includegraphics[width=\linewidth]{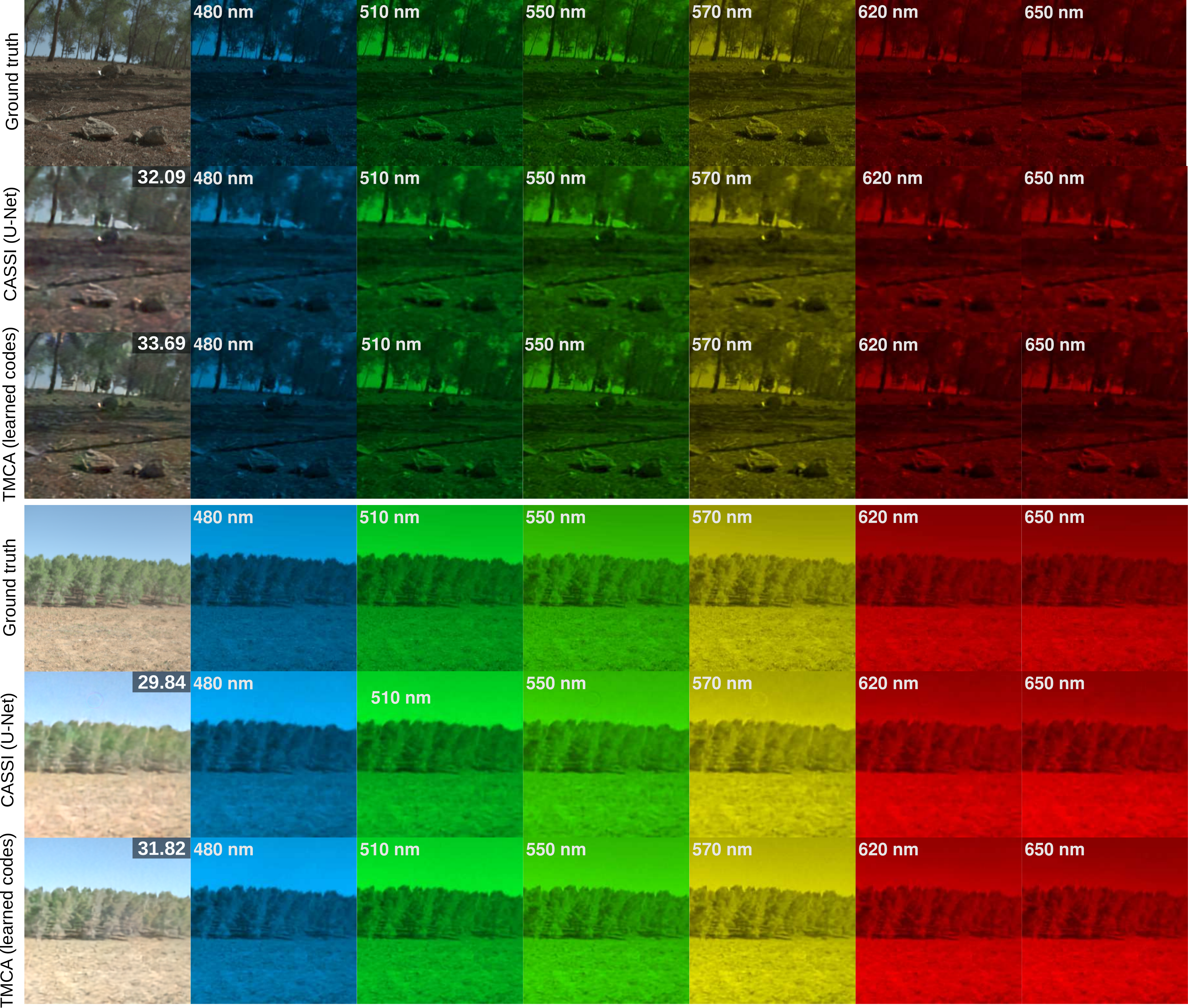}
    \caption{Results of our compressive spectral imaging reconstruction on two synthetic images of the ICVL 1 Dataset showing 6 different spectral bands (out of $L=12$).}
    \label{fig:image_spectral_bands}
\end{figure*}

\paragraph{Spectral signatures} We show qualitative and quantitative comparisons of the full spectral signatures ($L=12)$ for two different points taken in a randomly sampled image of the ICVL 1 dataset in Figure~\ref{fig:spectral_signatures} and show qualitatively $L=6$ bands for two other randomly sampled images in Figure~\ref{fig:image_spectral_bands}.
%
%
\begin{figure}[t!]
    \centering
    \includegraphics[width=\linewidth]{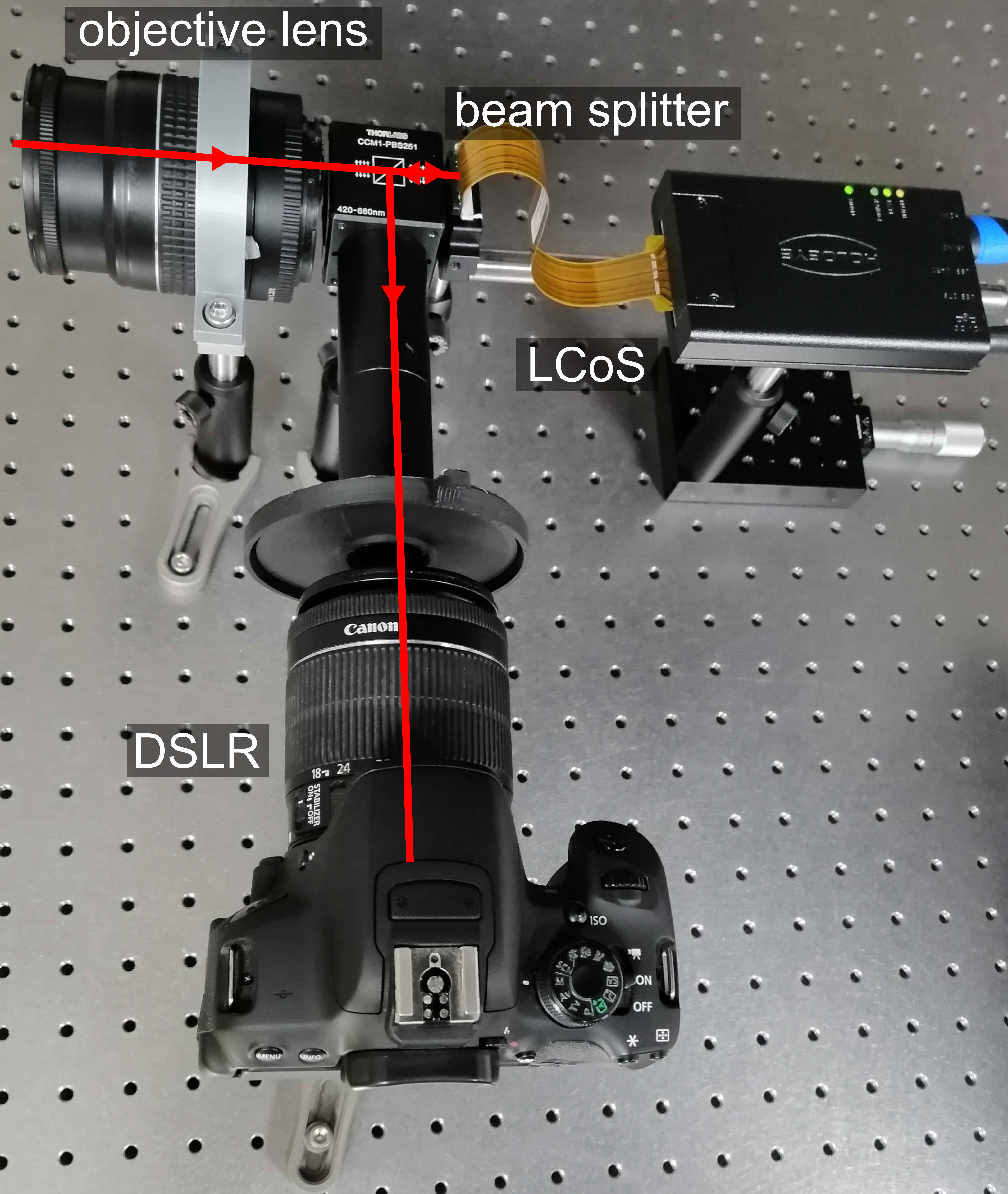}
    \caption{A photograph of the optical setup used in our compressive light field imaging experiments. The light path is shown in red and the various components are labelled (detailed components' description is found in the main paper).}
    \label{fig:clfp_optical_setup}
\end{figure}
\subsection{Compressive Light Field Imaging}
\paragraph*{Training details:} We learn our end-to-end model for compressive light field imaging by using the aggregate dataset described in the main paper. For, this experiment we aim to recover light fields with $5\times5$ angular views and resolution of $480\times270$ to match with the spatial resolution of the experimental setup. We also set the number of time slots in the TMCA encoder to $K=8$. We use randomly cropped patches of those images of spatial size $11\times11$ for training, the decoder is the deep spatial-angular convolutional sub-network proposed in \cite{guo2020deep} which is trained for 500 epochs using ADAM optimizer. Similar to the spectral application, we applied a learning rate decay of factor $0.5$ every $150$ epochs with an initial rate of $0.0001$. We display the learned coded apertures and shutter function in our pipeline in Figure~\ref{fig:coded_learned_lf}. We use a 2D array of sub-images for visualization of the TMCA where each sub-image represents the response of the equivalent coded aperture to all rays arriving at one point on the coded aperture from all points on the aperture plane.

\begin{figure*}[t!]
    \centering
    \includegraphics[width=\linewidth]{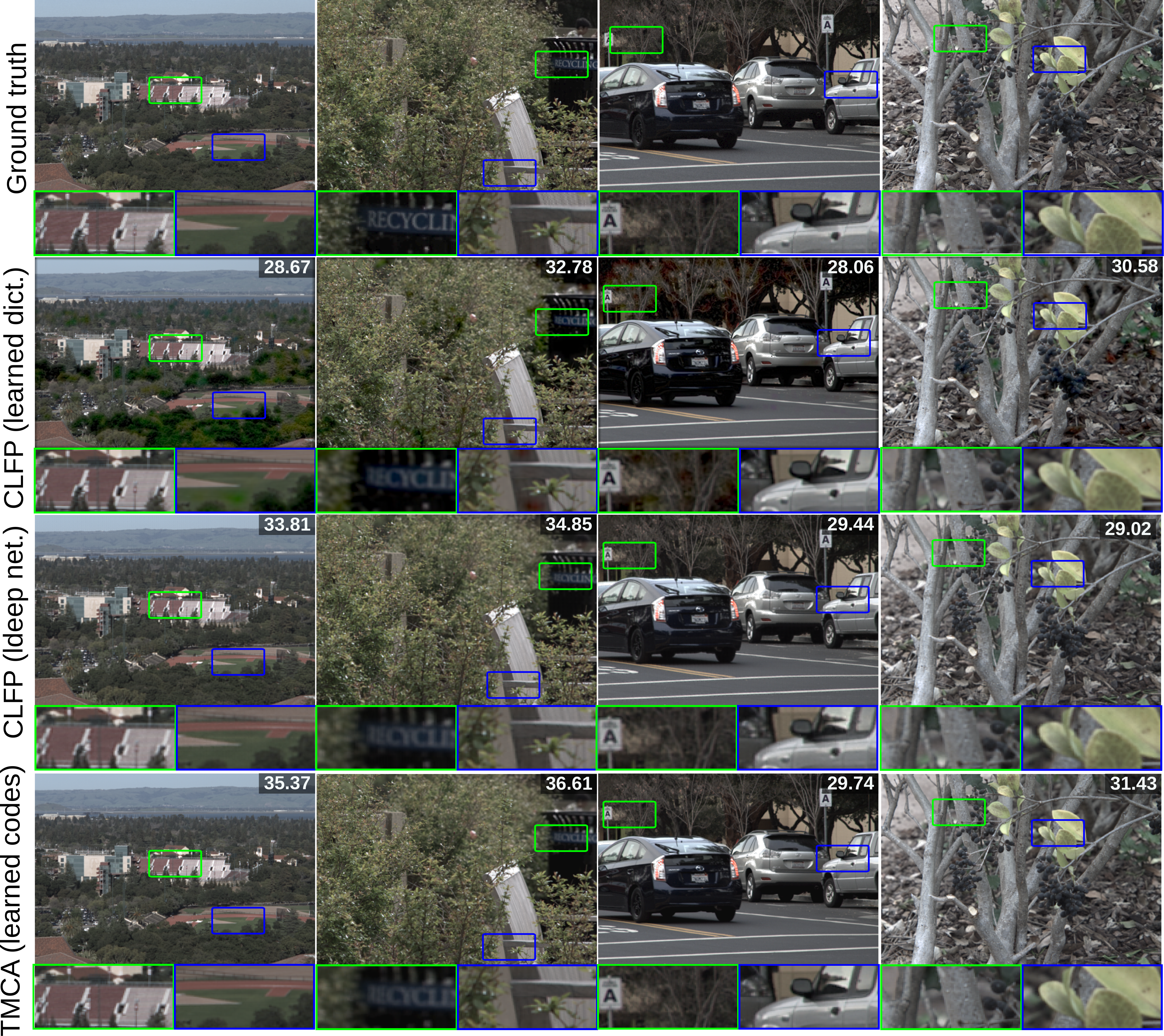}
    \caption{Additional light field results comparing our codification with TMCA using the deep network \cite{guo2020deep} as a decoder against (fourth row) the CLFP baselines with a sparse dictionnary coding method as a decoder (second row) and the same codification also using the deep network architecture from \cite{guo2020deep} (third row). Numbers in the top right corner indicate PSNR compared to the ground truth (first row).}
    \label{fig:LF_results}
\end{figure*}

\begin{figure*}[t!]
    \centering
    \includegraphics[width=0.85\linewidth]{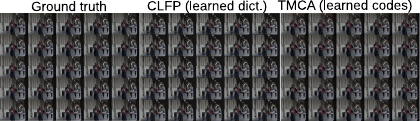}
    \caption{The $5\times5$ angular views reconstructed from a randomly sampled light field of the Lytro dataset comparing our method with the CLFP baseline.}
    \label{fig:LF_results_two}
\end{figure*}

\paragraph{Additional qualitative results} We show qualitative results for the central view of four additional reconstructed light fields in Figure~\ref{fig:LF_results} and show the $5\times5$ reconstructed angular views of yet another light field from the Lytro dataset in Figure~\ref{fig:LF_results_two}. 

\paragraph{Rational behind our baselines}
The rational behind our baselines for compressive light field imaging is the same as for hyperspectral imaging with baselines a) using the traditional codification and an ADMM decoder, b) swaping the ADMM  for a U-Net showing the U-Net alone does not explain the better results obtained using c) our TMCA codification or d) optimizing the codification with it. 
%
\section{Optical setups}
\paragraph{Compressive light field imaging} The setup consists of an objective lens projecting the image on a LCoS imaged with a DSLR (equipped with its objective lens) through a beamsplitter. The component details are given in the main paper. A photography of the setup used in our experiments in shown in Figure~\ref{fig:clfp_optical_setup}.

\begin{figure}[t!]
    \centering
    \includegraphics[width=\linewidth]{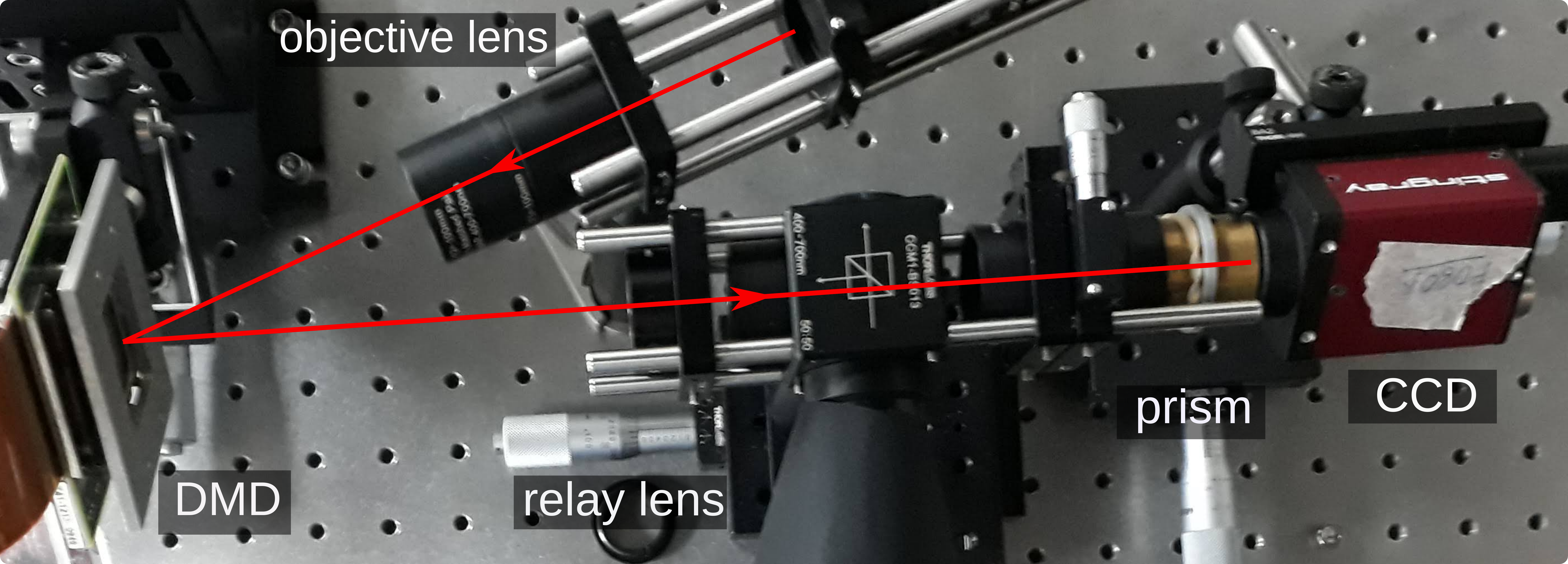}
    \caption{A photograph of the optical setup implementing our hyperspectral compressive imaging system. The light path is shown in red and the various components are labelled (a detailed components' description is found in the main paper.)}
    \label{fig:hyperspectral_optical_setup}
\end{figure}

\paragraph{Compressive hyperspectral imaging}
The setup consists of an objective lens projecting the image on a DMD imaged with a monochromatic CCD through a relay lens and prism dispersing light. The component details are given in the main paper. A photography of the setup used in our experiments in shown in Figure~\ref{fig:hyperspectral_optical_setup}.
%
%
%

\clearpage
%

\end{document}